# Recycling failed photoelectrons via tertiary photoemission

M. Matzelle[1,5], Wei-Chi Chiu[1,5], Caiyun Hong[2,5], Barun Ghosh[1], Pengxu Ran[2], R. S. Markiewicz[1], B. Barbiellini[4,1], Changxi Zheng[2], Sheng Li[3], Rui-Hua He[2†] & Arun Bansil[1†]

[1]Physics Department, Northeastern University, Boston, Massachusetts 02115, USA
[2]Research Center for Industries of the Future & Key Laboratory for Quantum Materials of Zhejiang Province, Department of Physics, Westlake University, Hangzhou, Zhejiang 310024, China
[3]Zhejiang Institute of Photoelectronics & Zhejiang Institute for Advanced Light Source, Zhejiang Normal University, Jinhua, Zhejiang 321004, China
[4]Department of Physics, School of Engineering Science, LUT University, Lappeenranta, Finland
[5]These authors contributed equally: M. Matzelle, Wei-Chi Chiu, Caiyun Hong.
†e-mail: ruihuahe@alumni.stanford.edu, ar.bansil@northeastern.edu

**A key insight of Einstein's theory of the photoelectric effect[1] is that a minimum energy is required for photoexcited electrons to escape from a material. For the past century it has been assumed that photoexcited electrons of lower energies make no contribution to the photoemission spectrum. Here we demonstrate the conceptual possibility that the energy of these 'failed' photoelectrons — primary or secondary — can be partially recycled to generate new 'tertiary' electrons of energy sufficient to escape. Such a 'recycling' step goes beyond the traditional three steps of the photoemission process (excitation, transport, and escape)[2,3] and, as we illustrate, it can be realized through a novel Auger[4] mechanism that involves three distinct minority electronic states in the material. We develop a phenomenological three-band model to treat this mechanism within a revised four-step framework for photoemission, which contains robust features of linewidth narrowing and population inversion under strong excitation, reminiscent of the lasing phenomena[5]. We show that the conditions for this recycling mechanism are likely satisfied in many quantum materials with multiple flat bands properly located away from the Fermi level, and elaborate on the representative case of $SrTiO_3$ among other promising candidates. We further discuss how this mechanism can explain the recent observation of anomalous intense coherent photoemission from a $SrTiO_3$ surface[6], and predict its manifestations in related experiments, including the 'forbidden' case of photoemission with photon energies lower than the work function. Our study calls for paradigm shifts across a range of fundamental and applied research fields, especially in the areas of photoemission, photocathodes, and flat-band**

**materials.**

Although Einstein received the Nobel Prize for his quantum theory of photoemission[1] published in 1905, it took over five decades before a comprehensive understanding of the photoemission process emerged[2,3]. In contrast to the then long-held belief that the process involves two steps (photoexcitation and escape) of a surface nature, Spicer proposed treating photoemission more generally as a bulk process. Justification for such a treatment was the recognition that electron-electron scattering which limits the electron escape depth to a few angstroms in metals, can be effectively suppressed within certain energy regions in the semiconductors called 'magic windows' (Fig. 1a), resulting in long lifetimes of the minority (primary and secondary) carriers. This necessitated the addition of a transport step that allows migration of electrons to the surface from a much larger volume of the material than previously thought. The resulting (three-step) model[2,3] turned out to be a cornerstone in understanding the photoemission process in solids and paved the way for later theoretical developments[7-11]. It provided crucial guidance for the development of practical photocathodes and photoemission-based instrumentation that underlie myriad technological advancements and scientific discoveries.

Six decades have passed since Spicer's original insight, and the fields of photoemission and photocathodes are perhaps ripe for another major transformation. Current strategies for developing efficient photocathodes are centered around placing the vacuum level ($E_{vac}$) within the magic window and keeping $E_{vac}$ as low as possible [12]. This general approach is rooted in the longstanding notion that the 'high-energy' minority electrons lying above the magic window and the 'low-energy' minority electrons lying below $E_{vac}$ contribute little or not at all to the photoemission spectrum[3]. We will show, however, that this is not the case in certain materials, in which both the high-energy and low-energy electrons can play a central role in the photoemission process. The conceptual basis of this emerging new paradigm is that, although the low-energy electrons cannot escape on their own, their energy can still be partially recycled or recovered to create new high-energy electrons with assistance from other excited states in the material. This constitutes a distinct step that needs to be added to Spicer's three steps for a more complete description of the photoemission process. This new step, which we call the recycling step, could in principle be realized through various up-conversion mechanisms depending on the nature of the involved excited states. Here we illustrate this step with an example based on a novel Auger mechanism that involves both the high-energy and low-energy minority electrons together with minority holes in the valence band as the relevant excited states. Our analysis indicates that the recycling step can dominate the overall photoemission process under certain conditions, resulting in substantial enhancements in both the efficiency and coherence of photoemission beyond the magic window.

## New conceptualization

Fig. 1b illustrates our new conceptual scheme. Photoexcitation creates minority electrons deep in the conduction band and holes in the valence band. These energetic carriers lose energy through electron-phonon scattering as they thermalize, with electrons approaching the conduction band minimum (CBM), while holes rise toward the valence band maximum (VBM). Carriers located beyond the magic windows can generate secondary electrons and holes through electron-electron scattering. Some of these (primary and secondary) electrons and holes enter two groups of excited states, defined at energies $E_E > E_{vac}$ and $E_L$, which we denote as the emission or E-band and the lower or L-band, respectively (Fig. 1b). Since the E-band lies outside the magic window, the associated electrons can only migrate over short distances, so that a small fraction of these electrons reach and escape from the surface, while a majority of them continue to thermalize. In this traditional three-step picture, one expects a relatively weak photoemission signal from the states in the E-band compared to that from the states near $E_{vac}$.

We now consider another group of states at energy $E_U$ (in the upper or U-band in Fig. 1b) and located $\Delta E$ below $E_{vac}$, as a host for minority (primary or secondary) electrons that enter below $E_{vac}$, including those that descend from the E-band. As these bands are located inside the magic windows for electrons and holes, respectively, their minority carriers possess long enough lifetimes for their (radiative or non-radiative) recombination to take place. For carriers that undergo Auger (non-radiative) recombination, the transition energy $E_U - E_L$ is passed onto other electrons, including the majority electrons in the lower-lying valence band. Some of these electrons in the vicinity of the intermediate or I-band at energy $E_I$ (Fig. 1b) are excited to the E-band, with the transition energy $E_E - E_I = E_U - E_L$, which we will refer to as the coincidence condition hereafter.

The aforementioned Auger transitions differ inherently from the traditional Auger processes[4]. Here the minority (U-band) electrons, rather than the majority electrons, initially fill the minority (L-band) holes. This is followed by the promotion of majority (I-band) electrons into the discrete (E-band) states rather than into the continuum. This new type of Auger process is second-order in the minority carrier density (see Methods). Therefore, for this process to yield a significant photoemission intensity, a high density-of-states (DOS) is required for both the U- and L-bands that serve to collect thermalized minority carriers. This novel Auger process can also be looked upon as a form of electron-electron scattering, in which a majority (I-band) electron is scattered by a minority (U-band) electron to fill a minority (L-band) hole. The excess energy is then released by promoting the scattered U-band electron to the E-band (Fig. 1b). Such an up-conversion is not possible via the type of electron-electron scattering traditionally involved in photoemission, where the original minority electron always loses (not gains) energy and the majority electron scatters to fill a majority (not minority) hole, yielding a secondary electron with energy always lower than that of the original minority electron[3]. To highlight its unique provenance, we will refer to the minority E-band electron generated through the novel Auger process as a tertiary electron.

Once these tertiary electrons enter the E-band, they merge with the short-lived primary and secondary electrons therein (Fig. 1b), effectively increasing the overall lifetime of the E-band electrons. As a result, additional electrons lying deeper in the bulk can potentially reach and escape from the surface, leading to an enhanced photoemission from the E-band. At the core of this process lie the novel Auger transitions, which partially collect failed photoelectrons and minority holes as the input and recycle their energies for the emission of tertiary electrons as output. This Auger-assisted tertiary photoemission (AATP) process thus plays the role of the recycling step, which we propose adding to the three traditional steps of photoemission.

## Theoretical modelling

We have developed a phenomenological model which captures the essential features of our revised (four-step) framework for photoemission. This minimal model consists of three coupled rate equations that describe the respective time evolutions of the minority sheet carrier densities, $n_i$ ($i = E, U, L$, hereafter), for the three bands [electrons in E- ($n_E$) and U-band ($n_U$) and holes in L-band ($p_L$)].

$$\frac{dn_E}{dt} = \alpha_E I_0 - \beta_E n_E + \gamma_E n_U p_L \tag{1}$$

$$\frac{dn_U}{dt} = \alpha_U I_0 - \beta_U n_U - \gamma_U n_U p_L + \beta n_E \tag{2}$$

$$\frac{dp_L}{dt} = \alpha_L I_0 - \beta_L p_L - \gamma_L n_U p_L \tag{3}$$

Here, the $\alpha_i$-terms are contributions of photoexcitation and thermalization processes to minority carrier inflows, all proportional to the excitation intensity $I_0$ with coefficients $\alpha_i$. The $\beta_i$-terms are contributions to minority carrier outflows due to scattering with defects, phonons, majority electrons, etc., all proportional to the densities and inverse lifetimes $\beta_i$ of minority carriers. The $\beta$-term refers to the carrier inflow to the U-band as a fraction of the outflow from the E-band, with the coefficient $\beta < \beta_E$. The $\gamma_i$-terms are contributions of the AATP process to carrier inflows to the E-band and outflows from the U- and L-bands, all proportional to the product of $n_U$ and $p_L$ with coefficients $\gamma_i$.

For alternative routes to recycling, other terms could be considered such as those representing photoexcitation of minority electrons or scattering between two minority electrons, etc. However, in focusing on the AATP process, we restrict ourselves to the simplified model of Eqs. 1-3. Note that the variables and parameters presented here and subsequently discussed are written in a dimensionless form. See Supplementary Information for discussion of their dimensionful versions and experimental relevance.

Eqs. 1-3 with physically constrained model parameters can be solved analytically for $n_i$ in the steady state (see Methods). The representative set of solutions shown in Fig. 2a as a function of $I_0$ highlight the following generic aspects of all steady-state solutions. The $I_0$-dependence of $n_E$ can be divided into three regions: when $I_0$ is small, $n_E$ is equal to $n_E^0$, the $n_E$ value calculated with all $\gamma_i$ set to zero, which is proportional to $I_0$ (i.e., a straight line with unit slope in the log-log plot); as $I_0$ increases, $n_E$ becomes larger than $n_E^0$, and shows a superlinear $I_0$-dependence; when $I_0$ is sufficiently large, the linear $I_0$-dependence is recovered for $n_E$ but with a much larger slope (Fig. 2a and inset). As a result, $n_E$ can become larger than one or both of $n_U$ and $p_L$, depending on the values of the model parameters (Extended Data Fig. 1), although it initially grows the slowest among the three densities. These results indicate that the AATP process can increase the minority electron population in the E-band, and that this AATP-induced population increase grows with $I_0$. Specifically, for $n_E > n_U$ in the large-$I_0$ region, the Maxwell-Boltzmann-like distribution of the thermalized electrons is violated for which a monotonically increasing density toward the CBM with $n_E < n_U$ would be expected[52]. This indicates that the AATP process can lead to a population inversion of the minority electrons between the E- and U-bands.

The effective inverse lifetimes of minority carriers, $\beta_{ei}$, can be obtained by studying the time-dependent decay of $n_i$ with $I_0$ set to zero at time $t = 0$ after the system reaches a steady state for any given $I_0$ (see Methods). In Fig. 2b, the calculated $\beta_{ei}$ are shown as a function of $I_0$ with the same set of model parameters as those used in Fig. 2a. Interestingly, the $I_0$-dependence of $\beta_{eE}$ can be divided into three regions corresponding to those in the case of $n_E$: in the small-$I_0$ region, $\beta_{eE}$ is equal to $\beta_{eE}^0$ ($=\beta_E$), the $\beta_{eE}$ value calculated with all $\gamma_i$ set to zero, which is independent of $I_0$; in the intermediate-$I_0$ region, $\beta_{eE}$ becomes smaller than $\beta_{eE}^0$ and exhibits a pronounced decrease with increasing $I_0$; in the large-$I_0$ region, $\beta_{eE}$ saturates to an $I_0$-dependent level that, depending on the values of model parameters, can be lower than $\beta_{eE}^0$ by orders of magnitude. As for $\beta_{eU}$ and $\beta_{eL}$, they start out to be smaller than $\beta_{eE}$ (with $\beta_E > \beta_U, \beta_L$) for small $I_0$. But, contrary to $\beta_{eE}$, both show a systematic increase with increasing $I_0$, and eventually become larger than $\beta_{eE}$. These results indicate that the AATP process can cause an effective lifetime enhancement of the minority electrons in the E-band. The AATP-induced lifetime enhancement grows with $I_0$, which can lead to an effective lifetime inversion of the minority carriers between the E- and U-/L-bands.

The above $I_0$-dependences of $n_E$ and $\beta_{eE}$ collectively reveal a picture that is robust against the chosen values of the model parameters (Extended Data Fig. 1) and thus generic to our three-band model. This picture depicts a crossover in the intermediate-$I_0$ region from the dominance by the regular photoemission process at lower $I_0$ to dominance by the AATP process at higher $I_0$. The crossover region is characterized by a superlinear growth of the population and a substantial enhancement of the effective lifetime of the minority electrons in the E-band. Because of their respective dominance by a single process, the low and high-$I_0$ regions are both associated with fixed ($I_0$-independent) population growth rates and lifetimes, despite their possible differences by orders of magnitude in the two regions.

## Materials realization

The conditions required for the recycling step in the form of the AATP process are satisfied in SrTiO$_3$. The calculated DOS based on density functional theory (DFT; see Methods) in Fig. 3a features a pair of van Hove singularities located at around 1.8 eV (Band 1) and -6.1 eV (Band 2) relative to the Fermi level ($E_F$). A subdominant feature in the valence band region is a relatively broad peak centered around -4.3 eV (Band 3). The bulk band structure (Fig. 3b) is seen to contain a parabolic electron-like band with its bottom at ~3.6 eV at the Γ point (Band 4), although this band is not visible in the DOS (Fig. 3a). Interestingly, Bands 1 and 4 are separated by ~1.8 eV, same as Bands 2 and 3. Note that the overall placement of these four band features in SrTiO$_3$ is much like that of the four bands depicted schematically in Fig. 1b. One can make clear assignments to these bands as labelled in Fig. 3a. Bands 1-3 are located within the magic windows for electrons and holes (orange-shaded regions), while Band 4 is located outside and above $E_{vac}$ (the green arrows; Fig. 3a-3b). This will allow the minority carriers to accumulate in high density and have long lifetimes in Bands 1 and 2, which are prerequisites for an effective AATP from Band 4. The coincidence condition, $E_E - E_I = E_U - E_L$, required to trigger the AATP process is also satisfied, which is in fact independent of the bandgap size.

The recycling mechanism can now be activated. Whether it can make a significant contribution to the overall photoemission process hinges on the probability of occurrence of the (non-radiative) novel Auger processes. A high probability would necessitate an effective suppression of the radiative recombination of the long-lived minority U-band electrons and L-band holes. We demonstrate how the latter likely takes place in SrTiO$_3$ by first noting the following aspects of its band structure: (1) As shown in Fig. 3e-3f, the U-band has little dispersion along the edges of the $k_x = 0$, $k_y = 0$ and $k_z = 0$ cross-sections of the simple cubic Brillouin zone ($X - M - Y$ and equivalent symmetry directions); (2) The L-band is two-dimensionally flat over these cross-sections (Fig. 3h-3i); (3) Both bands are derived from distinct atomic orbitals (Extended Data Fig. 2). For example, along $X - M$, the U-band mainly consists of Ti $3d_{xz}$ orbitals with a small admixture of O(3) $2p_z$ orbitals, while the L-band mainly consists of O(1) and O(2) $2p_z$, with a small admixture of Ti $3d_{yz}$. Their associated orbital wavefunctions are shown in Fig. 3g and 3j; (4) The orbitals involved in the U-band have virtually no overlap with those in the L-band. As illustrated in Fig. 3g and 3j for the momentum points along $X - M$, the Ti $3d_{xz}$ orbitals of U-band are orthogonal to Ti $3d_{yz}$ of L-band; the O $2p_z$ orbitals are located at different sites in both bands; and Ti 3d orbitals of one band do not overlap with O $2p_z$ of the other.

A necessary consequence of these band-structure features is a strong suppression in the (direct) dipole transitions between the states in the U- and L-bands. Consistent with this, the calculated joint DOS for direct transitions shows a distinct peak at 7.9 eV (=$E_U - E_L$), whereas this peak is absent in the calculated imaginary part of the dielectric function (Fig. 3d), which is proportional to the joint DOS with dipole-transition matrix element in the prefactor[13]. As a result, a major portion of the long-lived minority carriers in the U- and L-bands is expected to recombine non-radiatively and generate novel Auger transitions with high probability, once these bands are populated with minority carriers.

**Experimental connections**

Observation of anomalous intense coherent photoemission, which cannot be explained within the traditional three-step framework, has been reported recently from a reconstructed surface of SrTiO$_3$(001)[6]. In Methods, we discuss how the recycling mechanism can explain this observation and, based on our DFT and model calculations, provide guidance for further verification of this interpretation. Also, we predict below a new phenomenon that is uniquely enabled by this novel photoemission mechanism.

The common wisdom is that photoemission from a solid requires $hv$ to be greater than its surface work function $\phi = E_{vac} - E_F$. This also applies to semiconductors where $E_F$ is pinned within the bandgap due to the presence of in-gap states. As noted in Methods, $\phi_{AATP} = \max(E_U, |E_L|)$, the excitation threshold for the AATP process, is only determined by the energy positions of the U- and L-bands relative to $E_F$ and is independent of $\phi$. This suggests that for $\phi_{AATP} < \phi$, we will have the unusual situation in which purely tertiary photoemission will occur for $hv < \phi$. Specifically, in a material with $\phi_{AATP}$ as small as $(\phi_{AATP})_{\min}$ [= $(E_U - E_L)/2$], this unusual state of photoemission can be realized for any $\phi > (\phi_{AATP})_{\min}$, but only when the position of $E_F$ is located such that $\Delta E$ (= $E_{vac} - E_U$; as labelled in Fig. 1b) meets the condition $0 < \Delta E/2 < \min[\phi - (\phi_{AATP})_{\min}, (E_E - E_U)/2]$. SrTiO$_3$ has a theoretical value of $(\phi_{AATP})_{\min} \sim 3.95$ eV (Fig. 3b) and a typical experimental value of $\phi \sim 4.3$ eV that can vary up to 4.8 eV[14]. It can therefore become a viable platform for realizing the predicted state of photoemission, provided that both $E_{vac}$ and $E_F$ can be tuned to lie within the required energy ranges (see the pink arrow for a workable $E_F$ in Fig. 3b). Detecting a photoemission peak at a fixed energy $E_E > \phi$ with varying excitation energy $h\nu < \phi$ would establish this novel phenomenon and, in turn, serve as further evidence in support of our AATP mechanism. The associated spectral characteristics and their $h\nu$ dependence are anticipated to be qualitatively distinct from those of multi-photon photoemission which may become significant under strong excitation.

## Materials predictions

We have carried out a high-throughput screening for band structures that satisfy the key conditions for an effective AATP mechanism in two major materials databases[15-20] (see Methods). A cohort of candidate materials have been identified that range from cubic oxide perovskites (having band structures similar to SrTiO$_3$, see e.g. Fig. 4), to perovskite derivatives (including a perovskite solar cell material, see e.g. Extended Data Fig. 3), to non-perovskites (including a Kagome-lattice compound with a topological flat band located near $E_F$, see e.g. Extended Data Fig. 4). Many of these are quantum materials with intertwined degrees of freedom that would provide different knobs to tune their photocathode performance via the AATP process. For instance, the coincidence condition can be shown to be susceptible to strain in SrTiO$_3$ (Extended Data Fig. 5) and WO$_3$ (Fig. 4e and Extended Data Fig. 6). This will open a potential pathway for realizing strain-tunable [21] or flexible [22] photocathodes by growing or transferring thin films of related materials onto suitable substrates. Strong effects of electric polarization on the bulk electronic structure[23,24], surface potential[24], and strain[25,26] might permit electrical control of photocathode performance in ferroelectric systems, such as BaTiO$_3$ (Fig. 4b) [27,28], strained KTaO$_3$ (Fig. 4c) [29], and strained SrTiO$_3$ (Fig. 3) [30-32]. The predicted half-metallic ground state[33] of properly strained SrVO$_3$ photocathodes could drive spin-selective AATP processes[34,35] to yield spin-polarized electron beams (Fig. 4f). Experimental exploration of the above materials predictions promises the discovery of new photocathode quantum materials with novel functionalities beyond the reach of conventional photocathode materials.

## Broader implications

Our phenomenological three-band model within the four-step photoemission framework appears formally related to the Lotka-Volterra model well-known for describing predator-prey type population dynamics of relevance to many scientific disciplines [36-39]. This model was recently applied to describe population inversion and photon emission in a three-level laser system[40]. We have identified population inversion and linewidth narrowing as two robust features of our three-band model, which are also the hallmarks of the lasing phenomena[5]. Further work, however, is needed to understand the rich physical content of our three-band model, its possible physical connections with other aforementioned cases, as well as the precise relationship between the observed intense coherent photoemission[6] from $SrTiO_3$ and lasing.

At the core of the AATP process lie two flat bands (U- and L-bands) which can accommodate high-density, long-lived minority conduction electrons and valence holes with an enhanced tendency towards Auger decay. Both flat bands thus serve as a recycling engine that can effectively convert the energies of failed photoelectrons and minority holes into the energies of new valence electrons. The associated tertiary photoemission spectra show very narrow energy distributions centered at discrete energy positions, reminiscent of those seen in the resonant Auger emissions from atomic gases associated with core-excited states at much higher energies[41]. The above picture provides a new perspective on the nature of the observed intense coherent photoemission[6] from $SrTiO_3$ as an emergent atomic-like phenomenon due to the presence of flat conduction and valence bands. We note that current flat-band research has mainly focused on systems in which flat bands lie in close vicinity of $E_F$ as potential drivers of correlated electron physics[18]. Our study extends flat-band research to a greatly expanded pool of materials with flat-bands located away from the Fermi energy and the associated atomic-like responses and exotic properties that have remained hidden in plain sight.

1. Einstein, A. On a heuristic viewpoint concerning the production and transformation of light. *Ann. Phys.* **322**, 132-148 (1905). https://doi.org/10.1002/andp.19053220607.
2. Spicer, W. E. Photoemissive, photoconductive, and optical absorption studies of alkali-antimony compounds. *Phys. Rev.* **112**, 114 (1958).
3. Spicer, W. E. & Herrera-Gomez, A. Modern theory and applications of photocathodes. *Proc. SPIE* **2022**, 18 (1993).
4. Burhop, E. H. S. *The Auger effect and other radiationless transitions* (Cambridge University Press, 2014).
5. Svelto, O. *Principles of lasers* (Springer, 2010).
6. Hong, C. et al. Anomalous intense coherent secondary photoemission from a perovskite oxide. *Nature* **617**, 493-498 (2023).
7. Schaich, W. L., Ashcroft, N. W. Model calculations in the theory of photoemission. *Phys. Rev. B* **3**, 2452 (1971).
8. Pendry, J. B. Theory of photoemission. *Surf. Sci.* **57**, 679-705 (1976).

9. Nolting, W., Braun, J., Borstel, G. & Borgiel, W. Calculated photoemission and many body effects in 3d-ferromagnets. *Phys. Scr.* **41**, 601 (1990).
10. Bansil, A. & Lindroos, M. Importance of matrix elements in the ARPES spectra of BISCO. *Phys. Rev. Lett.* **83**, 5154-5157 (1999).
11. Braun, J., Minár, J. & Ebert, H. Correlation, temperature and disorder: Recent developments in the one-step description of angle-resolved photoemission. *Phys. Rep.* **740**, 1-34 (2018).
12. Rao, T. & Dowell, D. H. *An engineering guide to photoinjectors* (CreateSpace Independent Publishing, 2013).
13. Gajdoš, M., Hummer, K., Kresse, G., Furthmüller, J. & Bechstedt, F. Linear optical properties in the projector-augmented wave methodology. *Phys. Rev. B* **73**, 045112 (2006).
14. Ma, T., Jacobs, R., Booske, J. & Morgan, D. Understanding the interplay of surface structure and work function in oxides: A case study on $SrTiO_3$. *APL Mater.* **8**, 071110 (2020).
15. Jain, A. et al. Commentary: The Materials Project: A materials genome approach to accelerating materials innovation. *APL Mater.* **1**, 011002 (2013).
16. Munro, J. M., Latimer, K., Horton, M. K., Dwaraknath, S. & Persson, K. A. An improved symmetry-based approach to reciprocal space path selection in band structure calculations. *NPJ Comput. Mater.* **6**, 112 (2020).
17. Bradlyn, B. et al. Topological quantum chemistry. *Nature* **547**, 298-305 (2017).
18. Regnault, N. et al. Catalogue of flat band stoichiometric materials. *Nature* **603**, 824-828 (2022).
19. Vergniory, M. G. et al. All topological bands of all non-magnetic stoichiometric materials. *Science* **376**, 6595 (2022).
20. Călugăru, D. et al. General construction and topological classification of crystalline flat bands. *Nat. Phys.* **18**, 185 (2022).
21. Dhole, S., Chen, A., Nie, W., Park, B. & Jia, Q. Strain engineering: a pathway for tunable functionalities of perovskite metal oxide films. *Nanomaterials* **12**, 835 (2022).
22. Lu, D. et al. Synthesis of freestanding single-crystal perovskite films and heterostructures by etching of sacrificial water-soluble layers. *Nat. Mater.* **15**, 1255-1260 (2016).
23. Oshime, N. et al. Skewed electronic band structure induced by electric polarization in ferroelectric $BaTiO_3$. *Sci. Rep.* **10**, 10702 (2020).
24. Barrett, N. et al. Influence of the ferroelectric polarization on the electronic structure of $BaTiO_3$ thin films. *Surf. Interface Anal.* **42**, 1690-1694 (2010).
25. Stanescu, D. et al. Electrostriction, electroresistance, and electromigration in epitaxial $BaTiO_3$-based heterostructures: Role of interfaces and electric Poling. *ACS Appl. Nano Mater.* **2**, 3556-3569 (2019).
26. Zhang, J. et al. Large field-induced strains in a lead-free piezoelectric material. *Nat. Nanotech.* **6**, 98-102 (2011).
27. Acosta, M. et al. $BaTiO_3$-based piezoelectrics: Fundamentals, current status, and perspectives. *Applied Physics Reviews* **4**, 041305 (2017).

28. Kolodiazhnyi, T., Tachibana, M., Kawaji, H., Hwang, J. & Takayama-Muromachi, E. Persistence of ferroelectricity in $BaTiO_3$ through the insulator-metal transition. *Phys. Rev. Lett.* **104**, 147602 (2010).
29. Tyunina, M. et al. Evidence for strain-induced ferroelectric order in epitaxial thin-film $KTaO_3$. *Phys. Rev. Lett.* **104**, 227601 (2010).
30. Haeni, J. H. et al. Room-temperature ferroelectricity in strained $SrTiO_3$. *Nature* **430**, 758-761 (2004).
31. Xu, R. et al. Strain-induced room-temperature ferroelectricity in $SrTiO_3$ membranes. *Nat. Commun.* **11**, 3141 (2020).
32. Noël, P. et al. Non-volatile electric control of spin–charge conversion in a $SrTiO_3$ Rashba system. *Nature* **580**, 483-486 (2020).
33. Mahmood, Q., Ali, S. A., Hassan, M., Laref, A. First principles study of ferromagnetism, optical and thermoelectric behaviours of $AVO_3$ (A= Ca, Sr, Ba) perovskites. *Mater. Chem. Phys.* **211**, 428-437 (2018).
34. Yuan, J., Fritsche, L. & Noffke, J. Direct calculation of valence-band Auger emission: Spin polarization of Auger electrons from a potassium (110) surface. *Phys. Rev. B* **56**, 9942 (1997).
35. Kioupakis, E., Steiauf, D., Rinke, P., Delaney, K. T. & Van de Walle, C. G. First-principles calculations of indirect Auger recombination in nitride semiconductors. *Phys. Rev. B* **92**, 035207 (2015).
36. Abrams, P. A. The evolution of predator-prey interactions: Theory and evidence. *Annu. Rev. Ecol. Syst.* **31**, 79-105 (2000).
37. Hering, R. H. Oscillations in Lotka-Volterra systems of chemical reactions. *J. Math. Chem.* **5**, 197-202 (1990).
38. Ma Y.-A. & Qian, H. A thermodynamic theory of ecology: Helmholtz theorem for Lotka-Volterra equation, extended conservation law, and stochastic predator-prey dynamics. *Proc. R. Soc. A* **471**, 1-16 (2015).
39. Noonburg, V.W. A neural network modeled by an adaptive Lotka-Volterra system. *SIAM J. Appl. Math.* **49**, 1779-1792 (1989).
40. Aboites, V., Bravo-Avilés, J. F., García-López, J. H., Jaimes-Reategui, R. & Huerta-Cuellar, G. Interpretation and dynamics of the Lotka-Volterra model in the description of a three-level laser. *Photonics* **9**, 16 (2022).
41. Armen, G. B., Aksela, H., Åberg, T. & Aksela, S. The resonant Auger effect. *J. Phys. B: At. Mol. Opt. Phys.* **33**, R49-R92 (2000).

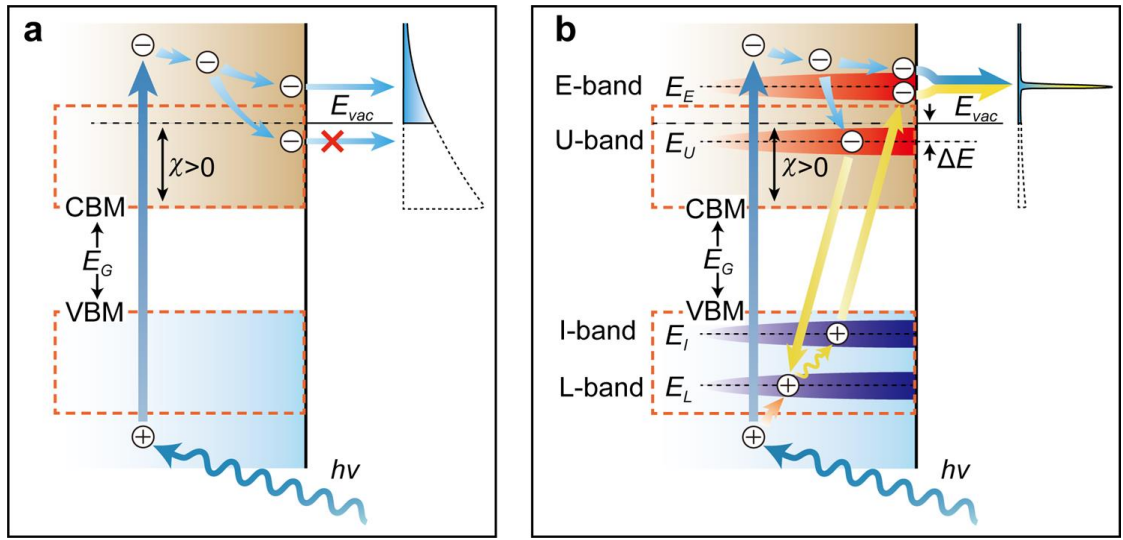

**Fig. 1| Schematic illustration of the traditional three-step framework and the proposed four-step framework for photoemission. a,** The traditional three-step framework in which the (regular) photoemission process is divided into three steps (photoexcitation, transport and escape). The photoexcited/thermalized minority electrons that reach the surface with energy lower than $E_{vac}$ (failed primary/secondary photoelectrons) cannot escape. **b,** The revised four-step framework in which a recycling step is added to the traditional three steps. The same band structure and positive electron affinity ($\chi > 0$) are assumed in **a** and **b** but with the presence of multiple distinct bands (as labelled) highlighted in **b**. On top of the regular photoemission process in **a**, the AATP process is highlighted with yellow arrows in **b**, which recycles the energy of failed photoelectrons for the emission of tertiary electrons from the E-band. Magic windows for minority carriers covering the range of the size of bandgap ($E_G$) right above/below the CBM/VBM are indicated by dashed rectangles. Note our definition of the magic window, which is different from its original version[3]. A schematic secondary photoemission spectrum (SPS) of a continuous (discrete) line shape is shown in the upper-right corner of **a** (**b**) in color shading with the portion cut off below $E_{vac}$ surrounded by dashed curves.

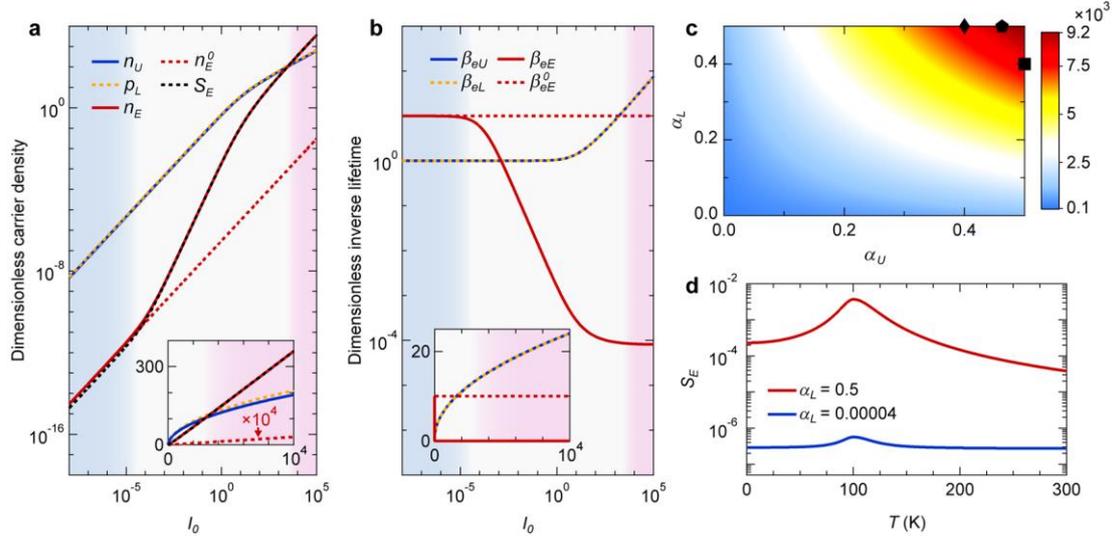

**Fig. 2 | Representative steady-state solutions of the phenomenological three-band model. a,** Carrier density $n_i$ and **b,** effective inverse lifetimes $\beta_{ei}$ for the E-, U- and L-bands as a function of $I_0$. Main panels (insets) are on logarithmic (linear) scales. **c,** Effective lifetime enhancement $L_E$ as a function of $\alpha_U$ and $\alpha_L$. **d,** Total number of the E-band electrons that can reach the surface and escape from a unit sample surface area $S_E$ as a function of temperature for two $\alpha_L$ values as labelled. In **a**, $n_E^0$ overlaps with $n_E$ in the low-$I_0$ region (blue) and deviates from it in the intermediate-$I_0$ (white) and high-$I_0$ (pink) regions. $S_E = \frac{n_E}{\beta_{eE} t_0}(1 - e^{-\beta_{eE} t_0})$, where $t_0 = d/v$, the diffusion time required for the E-band electrons of drift velocity $v$ to reach the surface from light absorption length $d$, see Methods for derivation. The difference between $S_E$ and $n_E$ decreases with $I_0$ as a result of decrease in $\beta_{eE}$. $n_E^0$ in the inset has been magnified by 10,000 times. In **b**, $\beta_{eE}^0$ overlaps with $\beta_{eE}$ in the same low-$I_0$ region. The overlap between $\beta_{eU}$ and $\beta_{eL}$ (also $n_U$ and $p_L$ in **a**) is accidental and it is due to the chosen values of the model parameters that fall into a specific parameter-space regime (see Methods for the definitions of the three regimes and Extended Data Fig. 1 for results in all three regimes with the corresponding $\alpha_U$ and $\alpha_L$ values marked in **c**). In **d**, A Lorentzian energy distribution of $\gamma_E = \frac{\gamma_{E0} \sigma^2}{(E(T) - E_{coin})^2 + \sigma^2}$ is assumed, centered at the energy position ($E_{coin}$) under the coincidence condition of a half-width $\sigma$ and amplitude $\gamma_{E0}$, where $E(T)$ is the temperature-dependent energy position of Peak 1 as approximated by the eyeguide in Extended Data Fig. 7e. The of model parameters used in all panels are: $\alpha_E = 3 \times 10^{-6}, \beta = 1, \beta_E = 10, \beta_U = \beta_L = 1, \gamma_U = \gamma_L = 0.12$. Additional parameters used are: $\gamma_E = 0.09$ in **a, b, c**; $\alpha_U = 0.4625$ in **a, b, d**; $\alpha_L = 0.5$ in **a, b**; $t_0 = 0.10$ in **a, d**; $I_0 = 1.43$ in **c, d**; $\gamma_{E0} = 0.09, E_{coin} = 4.086$ eV, $\sigma = 0.004$ eV in **d**. These values satisfy the physical constraints on various model parameters imposed by the conceptual picture illustrated in Fig. 1b (see Methods).

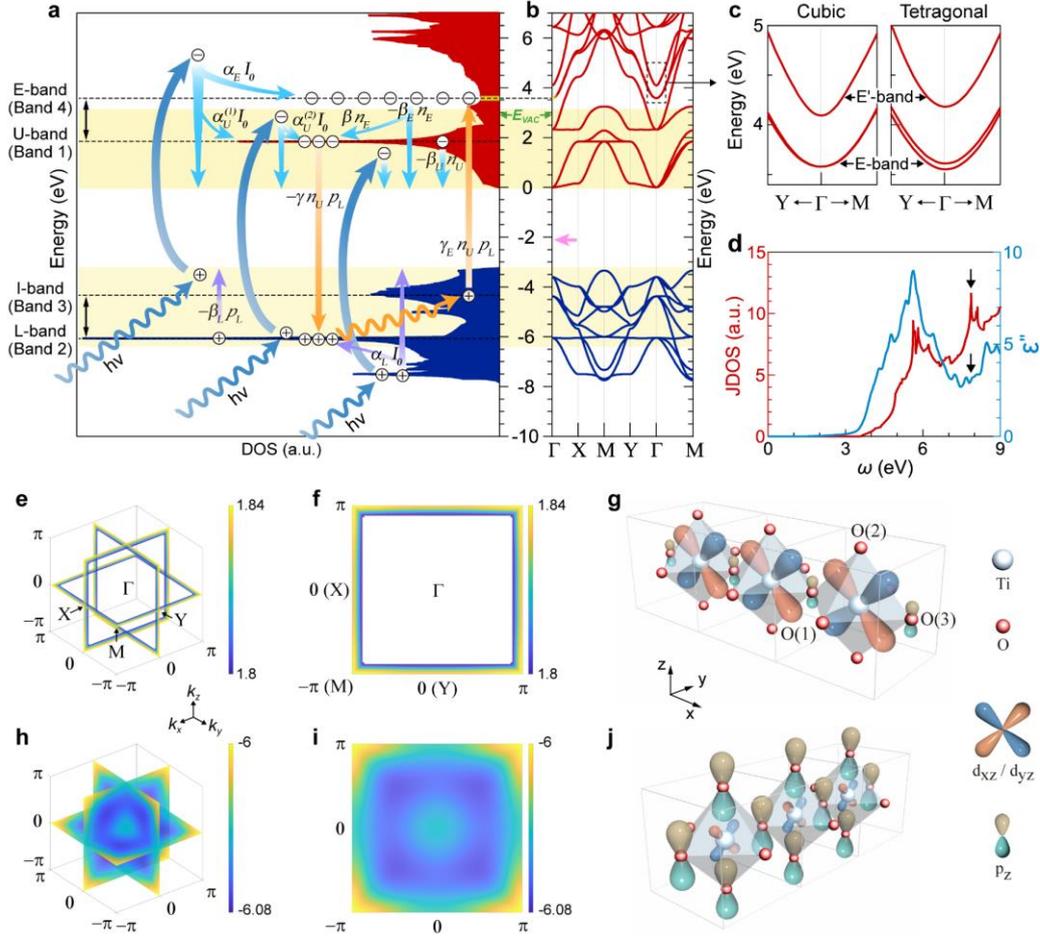

**Fig. 3 | Materials realization of the recycling mechanism in SrTiO$_3$. a,** DOS and **b,** band structure of bulk SrTiO$_3$ based on the DFT calculations (see Methods). Energy positions of the U-, L-, I- and E-bands (also denoted as Bands 1-4) are marked by the dashed lines. Two black arrows mark the equal energy separations between the E- and U-bands and between the I- and L-bands. The schematic illustration shows how the conceptual picture in Fig. 1b is realized in SrTiO$_3$. Green arrows mark the minimal value of $E_{vac}$ from the experiment[6] on SrTiO$_3$(001). Other colored arrows in **a** denote the photoexcitation step (blue), the minority electron (cyan) and hole (purple) thermalization/scattering processes in the transport step of the three-step framework, and the novel Auger processes (orange) in the recycling step of our proposed four-step framework. All terms in the phenomenological three-band model are labelled accordingly. $\alpha_U I_0$ has two thermalization contributions, $\alpha_U^{(1)} I_0$ and $\alpha_U^{(2)} I_0$, from the photoexcited final states above the E-band and those between the E- and U-bands, respectively. Magic windows for minority carriers (yellow shaded) are shown in **a, b**. **c,** Blowups of the E- and E'-bands for the cubic and tetragonal phases within the dashed rectangular region in **b**. **d,** Calculated joint DOS for direct transitions vs. the imaginary part of the dielectric function as a function of the energy transfer $\omega$. The arrows mark the (lack of) peak feature at $\omega = 7.9$ eV ($=E_U - E_L$). **e, f, g, (h, i, j,)** Dispersion contours shown in 3D and 2D Brillouin zones, and the schematic orbital wavefunction of the U-band (L-band) along $X - M$, respectively.

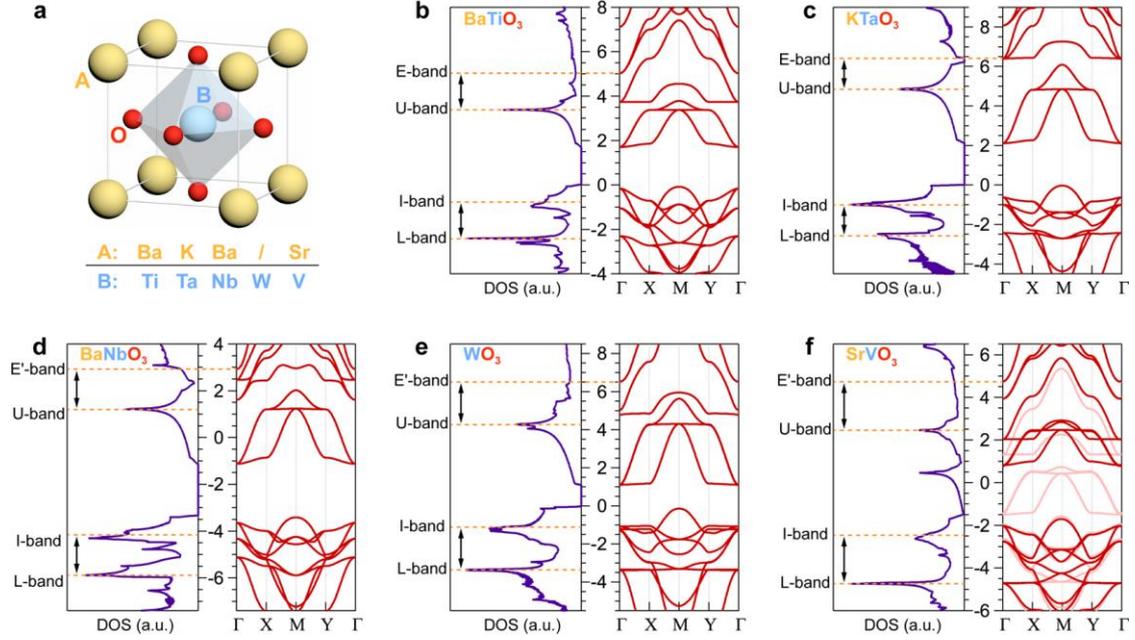

**Fig. 4 | Materials predictions of the cubic-oxide-perovskite candidates for the recycling mechanism. a,** Generic crystal structure of a cubic oxide perovskite with the chemical formula ABO$_3$. Various combinations of the A- and B-site elements are listed underneath for the transition-metal oxides (TMOs) shown in **b-f**. "/" stands for a missing site. **b-f,** Calculated DOSs and band structures of BaTiO$_3$ (d$^0$ non-strontium titanates), KTaO$_3$ (d$^0$ non-alkali-earth TMO), BaNbO$_3$ (d$^1$ TMO), WO$_3$ (d$^0$ simple-cubic TMO), and SVO$_3$ (d$^1$ half-metallic TMO). For each material, the E- (or E'-), U-, and L-bands are assigned to the DOS or band-structure features corresponding to those in SrTiO$_3$ (Fig. 3a). The I-band is obtained using the coincidence condition, and it is displaced from the L-band by the amount of the energy separation between the E- (or E'-) and U-bands, as marked by the arrows. Energy positions of the E- (or E'-), U-, I- and L-bands thus determined are marked by the orange dashed lines. Note that the I-band in each case lies in energy close to a local maximum in the DOS or the saddle point at X in the band structure, which indicates that the coincidence condition is met. Lattice parameters used in the bulk calculations here are taken from the database[15,16] and summarized in Methods. For SVO$_3$ in **f**, we obtain a ferromagnetic ground state with a spin-polarized band structure around $E_F$ (bands in the two spin channels shown in red and light red), consistent with an earlier report[33]. A spin-polarized U-band can drive spin-selective novel Auger transitions[34,35] in the AATP process. Note that the value of the lattice parameter from the database (3.901 Å) is 1.51 % larger than the experimental value for the bulk material that is non-magnetic[33]. An in-plane compressive strain of -0.63 % was applied to the database value of the lattice parameter (3.863 Å) in the calculations on WO$_3$ in **e** (see Methods and Extended Data Fig. 6). See Extended Data Figs. 3-4 for predictions of other candidate materials.

# Methods

## Phenomenological three-band model

### Physical assumptions and parameter constraints

We note the following points about our three-band model: 1) Defects, phonons, and majority carriers are involved in different processes of relevance to the model. But their densities are largely independent of time and $I_0$, and thus they can be incorporated into our definitions of the coefficients for the related processes. 2) $n_E$ is defined in the close vicinity of zero in-plane momentum ($k_\parallel$) associated with the final states measured in normal emission experiments, but $n_U$ and $p_L$ are defined after momentum integration over their respective energy bands. Due to the flat-band nature of the U- and L-bands, initial states at different momenta in both bands can participate in the novel Auger process with the same E-band final states. 3) Both $n_U$ and $p_L$ must be large to yield a significant AATP contribution to the photoemission process. The minority electrons (holes) in the U-band (L-band) can be generated in two ways, i.e., via direct photoexcitation to the U-band (L-band) and via thermalization of the states above the U-band (below the L-band) that are initially generated by photoexcitation. Since these are flat bands, many carriers can be collected in both ways at approximately the same energy but different momenta in each band, leading to a carrier density closely proportional to the DOS at the band energy. 4) $\alpha_i$ is related to the photoexcitation step, and thus it depends on $h\nu$. It also contains the contribution from the thermalization process in the transport step. A large $\alpha_i$ is generally expected for a band with a high DOS or a band that lies close to the band edges (CBM/VBM) so that it can receive many thermalized carriers, while the E-band around $k_\parallel$=0 is expected to have a very small $\alpha_E$ ($\ll \alpha_U, \alpha_L$). 5) $\beta_i$ is related to the scattering process in the transport step. Bands located inside the magic windows are expected to have smaller $\beta_i$ compared to those lying outside, hence $\beta_E > \beta_U, \beta_L$. 6) Recombination of electrons in the U-band and holes in the L-band will cause pair-wise reductions in their numbers, hence $\gamma_U = \gamma_L = \gamma$ (used in Fig. 3a). However, not all recombinations contribute to the upward novel Auger transition promoting electrons from the I-band to the E-band, hence $\gamma_E < \gamma_U, \gamma_L$.

### Steady-state solutions

In the steady state, as all densities reach equilibrium values, they no longer change as a function of time and their time derivatives vanish. Eqs. 1-3 in the main text can thus be rewritten as follows:

$$\beta_E n_E = \alpha_E I_0 + \gamma_E n_U p_L \quad (1)$$

$$\beta_U n_U = \alpha_U I_0 - \gamma_U n_U p_L + \beta n_E \quad (2)$$

$$\beta_L p_L = \alpha_L I_0 - \gamma_L n_U p_L. \tag{3}$$

Eqs. (2) and (3) can now be put on a more symmetric footing by defining $\alpha'_U = \alpha_U + \frac{\alpha_E \beta}{\beta_E}$ and $\gamma'_U = \gamma_U - \frac{\gamma_E \beta}{\beta_E}$ and substituting $n_E$ into Eq. (2):

$$\beta_U n_U = \alpha'_U I_0 - \gamma'_U n_U p_L \tag{4}$$

$$\beta_L p_L = \alpha_L I_0 - \gamma_L n_U p_L. \tag{5}$$

To decouple $n_U$ and $p_L$, we multiply Eq. (4) by $\gamma_L$ and Eq. (5) by $\gamma'_U$ and obtain:

$$\beta_L \gamma'_U p_L - \beta_U \gamma_L n_U = (\alpha_L \gamma'_U - \alpha'_U \gamma_L) I_0. \tag{6}$$

By defining $\alpha'_L = \frac{\beta_U \gamma_L}{\beta_L \gamma'_U}$ and $\gamma'_L = \frac{\alpha_L}{\beta_L} - \frac{\alpha'_U \gamma_L}{\beta_L \gamma'_U}$, Eq. (6) can be further reduced to

$$p_L = \alpha'_L n_U + \gamma'_L I_0. \tag{7}$$

By substituting Eq. (7) into Eq. (4), we can derive $\beta^+ n_U = \alpha'_U I_0 - \gamma^\dagger_L n_U^2$, where $\beta^+ = \beta_U + \gamma'_U \gamma'_L I_0$ and $\gamma^\dagger_L = \frac{\beta_U \gamma_L}{\beta_L}$, and solve this equation as:

$$n_U = \sqrt{\left(\frac{\beta^+}{2\gamma^\dagger_L}\right)^2 + \frac{\alpha'_U I_0}{\gamma^\dagger_L}} - \frac{\beta^+}{2\gamma^\dagger_L}. \tag{8}$$

Similarly, we can find the solution for $p_L$ as

$$p_L = \sqrt{\left(\frac{\beta^-}{2\gamma'_U}\right)^2 + \frac{\alpha^\dagger_L I_0}{\gamma'_U}} - \frac{\beta^-}{2\gamma'_U}, \tag{9}$$

where $\beta^- = \beta_U - \gamma'_U \gamma'_L I_0$ and $\alpha^\dagger_L = \frac{\beta_U \alpha_L}{\beta_L}$. By substituting Eqs. (8) and (9) into Eq. (1), we obtain

$$n_E = \frac{\alpha_E I_0}{\beta_E} + \frac{\gamma_E n_U p_L}{\beta_E}. \tag{10}$$

All steady-state minority carrier densities can now be determined from the fundamental parameters in Eqs. 1-3 in the main text. Note that the positive square root solutions are kept above for $n_U$ and $p_L$ as the carrier densities are expected to be greater than or equal to zero. From the definition of $\gamma'_U$, $\gamma'_U = \gamma_U - \frac{\gamma_E \beta}{\beta_E}$, $\gamma'_U$ is positive under our physical assumptions, $\beta < \beta_E$ and $\gamma_E < \gamma_U, \gamma_L$. In contrast, $\gamma'_L$ can be positive or negative if $\gamma'_U$ is positive, and it is positive definite if $\gamma'_U$ is negative. Note that the sign of $\gamma'_L$ in Eq. (7) indicates an important correlation between $n_U$ and $p_L$. In terms of the sign of $\gamma'_L$, three regimes ($\gamma'_L > 0$, $\gamma'_L = 0$ and $\gamma'_L < 0$) with unique $I_0$-dependent behaviors of carrier densities can be identified, see Extended Data Fig. 1 for examples.

Notably, beyond our focus in this work, if parts of the constraints used here ($\gamma_E < \gamma_U, \gamma_L$ and/or $\beta < \beta_E$) are relaxed, we can obtain $\gamma'_U < 0$. This case is discussed in Supplementary Information.

## Steady-state solutions in the $I_0 \to \infty$ limit

Due to the very large parameter space involved, we focus on the features of our model in the limiting case where the excitation intensity is very large ($I_0 \to \infty$).

For $\gamma'_L > 0$, we can then simplify the steady-state forms of $n_U$ and $p_L$ by taking the $I_0 \to \infty$ limit in Eqs. (8) and (9) as follows

$$\lim_{I_0 \to \infty} n_U = \frac{\alpha'_U}{\gamma'_U \gamma'_L} \tag{11}$$

$$\lim_{I_0 \to \infty} p_L = \gamma'_L I_0 \tag{12}$$

$$\lim_{I_0 \to \infty} n_E = \frac{I_0}{\beta_E}(\alpha_E + \Delta_U), \tag{13}$$

where $\Delta_U = \frac{\gamma_E \alpha'_U}{\gamma'_U}$. In this regime (Regime I), according to Eqs. (11)-(13), $n_U$ saturates to a constant that is independent of $I_0$, while both $p_L$ and $n_E$ are proportional to $I_0$.

For $\gamma'_L = 0$, $n_U$ is proportional to $p_L$ according to Eq. (7). In the $I_0 \to \infty$ limit, we obtain the following steady-state solutions.

$$\lim_{I_0 \to \infty} n_U = \sqrt{\frac{\alpha'_U I_0}{\beta}} \tag{14}$$

$$\lim_{I_0 \to \infty} p_L = \alpha'_L n_U \tag{15}$$

$$\lim_{I_0 \to \infty} n_E = \frac{I_0}{\beta_E}(\alpha_E + \Delta_=), \tag{16}$$

where $\Delta_= = \frac{\gamma_E \alpha'_L \alpha'_U}{\gamma'_U}$. In this regime (Regime II), both $n_U$ and $p_L$ are proportional to $\sqrt{I_0}$, while $n_E$ is proportional to $I_0$.

For $\gamma'_L < 0$, in the $I_0 \to \infty$ limit, we derive the following steady-state solutions.

$$\lim_{I_0 \to \infty} n_U = \frac{|\gamma'_L| I_0}{\alpha'_L} \tag{17}$$

$$\lim_{I_0 \to \infty} p_L = \frac{\alpha'_L \alpha_L}{\gamma_L |\gamma'_L|} \tag{18}$$

$$\lim_{I_0 \to \infty} n_E = \frac{I_0}{\beta_E}(\alpha_E + \Delta_L), \tag{19}$$

where $\Delta_L = \frac{\gamma_E \alpha_L}{\gamma_L}$. In this regime (Regime III), $p_L$ saturates to a constant that is independent of $I_0$, while both $n_U$ and $n_E$ are proportional to $I_0$.

In summary, in the large-$I_0$ limit, for $\gamma'_L > 0$ ($\gamma'_L < 0$), $n_U$ ($p_L$) saturates to a steady value that is independent of $I_0$, while $p_L$ ($n_U$) and $n_E$ are proportional to $I_0$. The regime where $\gamma'_L = 0$ is unique in that in this case $n_U$ is proportional to $p_L$ and neither of them saturates. These behaviors are clearly seen in the insets of Extended Data Fig. 1a-1c.

**Effective inverse lifetimes**

While Eqs. 1-3 in the main text describe the growth process under photoexcitation ($I_0 \neq 0$), we consider the decay process ($I_0 = 0$) of the carrier densities when the excitation is turned off (after it is turned on for some time) as follows,

$$\frac{dn_E}{dt} = -\beta_E n_E + \gamma_E n_U p_L \tag{20}$$

$$\frac{dn_U}{dt} = -\beta_U n_U - \gamma_U n_U p_L + \beta n_E \tag{21}$$

$$\frac{dp_L}{dt} = -\beta_L p_L - \gamma_L n_U p_L. \tag{22}$$

Here, we gain some insight into the decay rate equations by assuming that the carrier densities decay exponentially: $\frac{dn_E}{dt} = -\beta_{eE} n_E$ ; $\frac{dn_U}{dt} = -\beta_{eU} n_U$ ; $\frac{dp_L}{dt} = -\beta_{eL} p_L$, with effective inverse lifetimes (decay rates) $\beta_{ei}$. The Eqs. (20)-(22) thus can be written as

$$\beta_{eE} n_E = \beta_E n_E - \gamma_E n_U p_L \tag{23}$$

$$\beta_{eU} n_U = -\beta n_E + \beta_U n_U + \gamma_U n_U p_L \tag{24}$$

$$\beta_{eL} p_L = \beta_L p_L + \gamma_L n_U p_L. \tag{25}$$

We further consider the case where all relevant carrier densities reach steady state with the excitation on for a sufficiently long time, and then the excitation is turned off at time $t = 0$. The initial carrier densities at $t = 0$ for the decay process (excitation off) are thus equal the steady state values for the growth process (excitation on), and the effective inverse lifetimes $\beta_{ei}$ are the initial lifetimes at $t = 0$ in the decay process, or equivalently, the steady-state inverse lifetimes. The effective inverse lifetimes $\beta_{ei}$ of steady state thus can be expressed as a function of $I_0$ by combining Eqs. (1)-(3) with Eqs. (23)-(25):

$$\beta_{eE} = \frac{\alpha_E}{n_E} I_0 \tag{26}$$

$$\beta_{eU} = \frac{\alpha_U}{n_U} I_0 \tag{27}$$

$$\beta_{eL} = \frac{\alpha_L}{p_L} I_0, \tag{28}$$

where $n_E, n_U$ and $p_L$ are the steady-state solutions in Eq. (8)-(10). These $\beta_{ei}$ of the steady state as a function of $I_0$ are presented in Extended Data Fig. 1d-1f.

For $t > 0$, after the decay process has started, we need to consider the time-dependent behavior of the carrier densities $n_i(t)$, and we will have the associated time-dependent effective inverse lifetimes $\beta_{ei}(t, I_0)$. The full description of the time-dependent effective inverse lifetimes is presented in Supplementary Information.

**Effective lifetime enhancement**

One of the important features of our phenomenological model is the prediction of lifetime-enhanced minority electrons in the E-band due to the novel Auger feedback term $(\gamma_E n_U p_L)$. To quantify this effect, we define the effective lifetime enhancement for the E-band by comparing the effective inverse lifetime $\beta_{eE}$ to the intrinsic inverse lifetime $\beta_E$ as

$$L_E(I_0) = \frac{\beta_E}{\beta_{eE}(I_o)}. \tag{29}$$

Then, the effective lifetime enhancement for the steady state can be derived from Eqs. (10) and (26) as

$$L_E = \frac{\beta_E n_E}{\alpha_E I_0} = 1 + \frac{\beta_E \gamma_E}{\alpha_E I_0} n_U p_L. \tag{30}$$

We see that $L_E$ is proportional to the carrier density of $n_E$. This is illustrated as a function of $\alpha_U$ and $\alpha_L$ in Fig. 2c.

To gain further insight into the behavior of $L_E$, we consider the effective inverse lifetimes $\beta_{ei}$ with the steady-state solution in the $I_0 \to \infty$ limit, and discuss the three regimes separately ($\gamma'_L > 0, \gamma'_L = 0$ and $\gamma'_L < 0$).

In Regime I ($\gamma'_L > 0$), we can derive $\lim_{I_0 \to \infty} \beta_{eE} = \frac{\alpha_E}{(\alpha_E + \Delta_U)} \beta_E$ by plugging Eq. (13) into Eq. (26), so that

$$\lim_{I_0 \to \infty} L_E = 1 + \frac{\Delta_U}{\alpha_E}. \tag{31}$$

In Regime II ($\gamma'_L = 0$), by using Eq. (16), we can obtain $\lim_{I_0 \to \infty} \beta_{eE} = \frac{\alpha_E}{(\alpha_E + \Delta_=)} \beta_E$ and

$$\lim_{I_0 \to \infty} L_E = 1 + \frac{\Delta_=}{\alpha_E}. \tag{32}$$

In Regime III ($\gamma'_L < 0$), we obtain $\lim_{I_0 \to \infty} \beta_{eE} = \frac{\alpha_E}{(\alpha_E + \Delta_L)} \beta_E$ from Eqs. (19) and (26), yielding

$$\lim_{I_0 \to \infty} L_E = 1 + \frac{\Delta_L}{\alpha_E} . \tag{33}$$

**$T$- and $h\nu$-dependence of the model parameters**

Since currently $n_U$ or $p_L$ are not accessible experimentally, we are not in a position to provide a detailed understanding of the observed $T$- and $h\nu$-dependence of $n_E$. However, we can show that our model has adequate flexibility to describe both the strong growth of $n_E$ at low $T$ for $h\nu > \phi_{AATP}$, and the corresponding weak growth for $h\nu \cong \phi_{AATP}$ as implied by the recent experimental results (Extended Data Fig. 7d). Remarkably, this can be explained rather simply in terms of the $h\nu$-dependence of the $\alpha_L$ term and the $T$-dependence of the E-band energy, see Fig. 2d.

Due to the proximity of $h\nu$ =6.2 eV to $\phi_{AATP}$, set by the L-band energy, the $\alpha_L$ term (hence $p_L$) should be correspondingly small leading to a very small novel Auger enhancement (Fig. 2c). To model the difference between $h\nu = 6.2$ eV and 10.0 eV, we choose the parameter $\alpha_L$ to be $4 \times 10^{-5}$ and 0.5, respectively. The measured $T$-dependence of the E-band energy (Extended Data Fig. 7e) strongly affects the coincidence condition required for the novel Auger process, as explained below. However, to make the comparison, we must first establish the relationship between the experimentally observed E-band intensity and $n_E$.

*E-band electrons escaping per unit sample surface area:* The measured integrated intensity of the E-band photoemission, which is proportional to the total number of E-band electrons escaping per unit sample surface area $S_E$, is related to the minority carriers (sheet) density $n_E$ as follows. We assume that within the light absorption length $d$, the number of E-band minority electrons per unit volume is given by $n_d$. The probability that an electron from a certain depth $z$ can escape is given by $e^{-\beta_{eE}\frac{z}{v}}$, where $\beta_{eE}$ is the effective inverse lifetime as defined previously, and $v$ is the drift velocity of the E-band electrons.

To get the total electrons that escape $N_E$, we integrate $n_d$ multiplied by the probability of escaping from depth $z$ over the sample area $A$.

$$N_E = \iiint n_d \, e^{-\beta_{eE}\frac{z}{v}} dxdydz = n_d A \frac{v}{\beta_{eE}}\left(1 - e^{-\beta_{eE}\frac{d}{v}}\right). \tag{34}$$

By dividing both sides by the surface area $A$ we obtain

$$S_E = \frac{N_E}{A} = n_d \frac{v}{\beta_{eE}}\left(1 - e^{-\beta_{eE}\frac{d}{v}}\right) = \frac{n_E}{\beta_{eE} t_0}\left(1 - e^{-\beta_{eE} t_0}\right), \tag{35}$$

where in the final step we have used that $n_d d = n_E$ and defined $t_0 = d/v$.

*Novel Auger resonance:* The strong correlation between the peak intensity and the peak position in Extended Data Fig. 7d-7e indicates that the temperature dependence is likely connected to the deviation from satisfying the coincidence condition perfectly. At zero temperature, the coincidence condition for the novel Auger process is enforced by a Dirac delta term: $\delta(\epsilon_1 + \epsilon_2 - \epsilon_3 - \epsilon_4)$ with the $\epsilon_i$'s being the energies of the four involved energy levels. At finite temperatures, since the discrete energy levels can be shifted and broadened, we replace the Dirac delta can be replaced by a Lorentzian of the form: $Z = \frac{1}{\pi}\frac{\sigma}{(\Delta\epsilon)^2 + \sigma^2}$. Here $\Delta\epsilon$ is equal to the argument of the Dirac delta term with each energy level's respective energy shift and $\sigma$ is the related inverse lifetime.

We assume that other possible energy shifts with temperature are small compared to the E-band shift and use $\sigma$ as a fitting parameter. $\Delta\epsilon$ can then be expressed as $E(T) - E_{coin}$, where $E(T)$ is the center of E-band energy and $E_{coin}$ is the E-band energy center that fulfills the coincidence condition perfectly. To ensure that $\gamma_E$ reaches its maximum value when the coincidence condition is met, we define $\gamma_E(T) = \sigma\pi\gamma_{E0}Z$, which has a maximum $\gamma_{E0}$ at $E(T) = E_{coin}$. Note that $\gamma_{E0}$ is chosen to be consistent with our physical assumption that $\gamma_E(T) < \gamma_U$. Furthermore, we assume that the coincidence condition is met at 100 K, i.e. $E(100\,K) = E_{coin}$. $E(T)$ can now be directly obtained from the experiment by fitting the data points in Extended Data Fig. 7e to a simple power-law $T$-dependence, $E(T) = 4.1012 - 9.77 \times 10^{-7} \times T^{2.1}$, between 0 K and 100 K and to a linear $T$-dependence, $E(T) = 1.966 \times 10^{-4} \times (T - 100)$, between 100 K and 300 K.

## Band structure calculations for SrTiO$_3$

First-principles DFT-based calculations were carried out using the projector augmented-wave pseudopotentials[42] using the Vienna Ab initio Simulation Package (VASP)[43]. An energy cutoff of 520 eV was chosen for the plane-wave basis set and the Brillouin zone was sampled with a Γ-centered mesh of 15×15×15 points. The generalized gradient approximation (GGA) with the parameterization of Perdew, Burke and Ernzerhof (PBE)[44] was used for the exchange-correlation functional. All calculations were self-consisted with a total energy tolerance of 1×10$^{-7}$ eV. The lattice parameters as well as the atomic positions were relaxed by minimizing the forces with a tolerance of 8×10$^{-3}$ eV/Å. The relaxed lattice parameter of *a*= 3.941 Å for cubic SrTiO$_3$ agrees well with a previous study using the same method[45].

Additional bulk calculations were performed on the tetragonal phase with the lattice parameters, $a$= 5.563 Å and $c$= 7.900 Å and using the experimental twist angle of 1.4° after Ref. 45. Differences in the band structure from the cubic phase are shown in Extended Data Fig. 5a. Also, calculations on the cubic phase were repeated using a reduced lattice parameter of $a$= 3.900 Å to ascertain changes in the band structure, see in Extended Data Fig. 5d. Finally, using the strongly-constrained-and-appropriately-normed (SCAN) meta-GGA functional[46], we investigated the robustness of our results, see Extended Data Fig. 5b. Notably the SCAN-based relaxed lattice parameter, $a$= 3.908 Å is quite close to the experimental value of $a$= 3.900 Å.

The surface-state and DOS calculations were performed by employing a real-space tight-binding model Hamiltonian, which was obtained by using the VASP2WANNIER90 interface[47,48]. We included the s, p and d orbitals of Ti and Sr atoms and the s and p orbitals of the three O atoms to generate the Wannier functions. DOSs of the cubic and tetragonal phases were calculated by using 501×501×501 and 355×355×251 point meshes in their respective first Brillouin zones to make the momentum-space resolutions comparable in both calculations. The joint DOS for direct transitions was calculated with a 501×501×501 point mesh using, $\text{JDOS}(\omega) = \int \frac{d\mathbf{k}}{(2\pi)^3} \sum_{c,v} \delta(E_{c,\mathbf{k}} - E_{v,\mathbf{k}} - \omega)$, where $c$ and $v$ refer to for all conduction and valence bands, respectively. The surface and bulk band structures of the semi-infinite slab were calculated using the iterative Green's function method[49]. The DOS of semi-infinite slab was calculated using a 501 × 501 point mesh in the (001) surface Brillouin zone.

A typical ('scissor') operation[50] was applied to the calculated band structures using a rigid shift of all the conduction bands in energy such that the bandgap between the CBM at Γ and the VBM at $A$ is equal to the reported experimental value of 3.25 eV. The imaginary part of the dielectric function was calculated within the framework of the independent particle approximation[13].

## Experimental benchmarking for SrTiO$_3$

Three most remarkable findings of the recent SPS study[6] of SrTiO$_3$(001) are: a discrete lineshape of the SPS that features two peaks lying well above $E_{vac}$ (Finding 1; Extended Data Fig. 7a-7b); a substantial linewidth narrowing and increase of the integrated intensity of the SPS peaks at the photon energy $h\nu$ =10 eV in comparison with $h\nu$ =6.2 eV (Finding 2; Extended Data Fig. 7d and 7b), and a strong enhancement of the integrated intensity of the SPS peak upon lowering temperature (Finding 3; Extended Data Fig. 7c). All these findings can be reasonably captured by the proposed recycling mechanism as discussed below.

Notably, our calculations on SrTiO$_3$ find in the vicinity of the E-band another E′-band with a similar parabolic electron-like in-plane dispersion centered at Γ (Fig. 3b-3c). Both bands display an in-plane isotropic dispersion around the Γ − Z direction (Extended Data Fig. 8g-8h). The E- and E′-bands have slightly different effective masses at Γ, ~0.64 m$_e$ and ~0.46 m$_e$ (m$_e$ is the free electron mass), respectively. Their energy positions and effective masses vary with the out-of-plane momentum $k_z$ (Extended Data Fig. 8a-8b), with the exact values depending to some extent on the details of calculations (Extended Data Fig. 5b-5d). The two-fold degeneracy of the E-band at Γ in cubic SrTiO$_3$ is lifted in the tetragonal phase with the appearance of a small splitting (Fig. 3c and Extended Data Fig. 5a). Related to Finding 1 noted above on the experimental Peaks 1 and 2, the observed dispersions and energy ranges of these peaks are similar to those of the calculated E- and E′-bands (Extended Data Fig. 8a-8d). Their effective masses are also in reasonable accord with the calculated values (Extended Data Fig. 9). The extra peak (Peak 1′ in Extended Data Fig. 8d) observed to emerge very close to Peak 1 upon cooling below ~100 K[6], is reminiscent of the calculated E-band splitting in the tetragonal phase. With the above agreements between the experiment and DFT calculations in mind, we conclude that the experimental Peaks 1 and 2 can be reasonably ascribed to the calculated E- and E′-band, respectively.

Our theoretical modelling shows that for activating the AATP contribution to photoemission from the E-band, both $n_U$ and $p_L$ must be non-zero in the steady state. This can be realized by exciting the electrons (holes) in the vicinity of $E_F$ to the U-band (L-band) with a minimal amount of energy. The excitation threshold for the AATP process thus is: $\phi_{AATP} = \max(E_U, |E_L|)$, which is ~6.1 eV in SrTiO$_3$ (Fig. 3b). For $h\nu$ <6.1 eV, $p_L$ will be zero due to lack of minority hole inflow to the L-band caused by photoexcitation and thermalization ($\alpha_L$=0), so that AATP will not be activated and photoemission from the E-band will be negligible. For $h\nu \gtrsim 6.1$ eV, $p_L$ and $\alpha_L$ become non-zero but remain small due to the relatively low DOS of the initial states near $E_F$. But $n_U$ will not be small because the valence band of a much higher DOS can feed the U-band with minority electrons, hence $\alpha_U > \alpha_L$. As a result, the AATP contribution to photoemission becomes non-zero, with $p_L$ being a limiting factor.

As $h\nu$ increases towards ($E_U - E_L$)~7.9 eV (Fig. 3b), we expect both $p_L$ and $n_U$ to increase continuously, along with the AATP contribution to photoemission. The reason is that both $\alpha_U$ and $\alpha_L$ increase as more electrons (holes) closer to the L-band (U-band) are photoexcited and thermalized to the U-band (L-band). When $h\nu$ goes above 7.9 eV, thermalization of additional minority carriers from the energy regions above (below) the U-band (L-band) start to contribute to the inflows into the other band. The AATP contribution continues to increase but at a moderate rate because the DOS in those energy regions does not contain singular features (Fig. 3b). Increasing $h\nu$ from below 6.1 eV to above 7.9 eV thus leads to an increase in the AATP contribution via increases in $\alpha_U$ and $\alpha_L$. Within our phenomenological model, the effective lifetime enhancement, $L_E$ (= $\beta_E/\beta_{eE}$), is proportional to $n_E$ [Eq. (30)], while the total number of E-band electrons that can reach the surface and escape per unit sample area, $S_E$, is also proportional to $n_E$ [Eq. (35); Fig. 2a]. Accordingly, increases in $\alpha_L$ and $\alpha_U$ lead to increases in both $L_E$ and $S_E$, as seen in Fig. 2c for $L_E$ based on the same set of model parameters (other than $\alpha_L$ and $\alpha_U$) as those used in Fig. 2a-2b. Increases in $L_E$ and $S_E$ will be reflected as linewidth narrowing and increase of the integrated intensity of the E-band photoemission peak, respectively. Both these aspects revealed by our theoretical model are in accord with experimental Finding 2 noted above on the linewidth and integrated intensity of Peak 1 (Extended Data Fig. 7b-7c), suggestive of a substantially enhanced AATP contribution to photoemission for $h\nu \geq 10.0$ eV over that for $h\nu =6.2$ eV near its excitation threshold.

To maximize the AATP contribution to photoemission from the E-band, the coincidence condition ( $E_E - E_I = E_U - E_L$ ) must be met precisely. Any deviation from this condition will decrease the AATP contribution by decreasing $\gamma_E$. Given that the Peak 1 intensity was observed to maximize around 100 K in SrTiO$_3$ (Extended Data Fig. 7d), it is reasonable to assume that the coincidence condition is fulfilled almost perfectly for Peak 1 at 100 K, and that this is no longer the case at higher temperatures as Peak 1 moves away from its 100 K position (Extended Data Fig. 7e). We have approximated the temperature dependent AATP contribution by only using a $\gamma_E$ of a Lorentzian form, which peaks at 100 K and falls with temperature to mimic the effect of the observed Peak 1 movement (see the earlier subsection "$T$- and $h\nu$-dependence of the model parameters"). The resulting model calculations correctly capture salient aspects of the experimental Finding 3 noted above (Extended Data Fig. 7d). Specifically, the integrated intensity of Peak 1 observed at $h\nu =10.0$ eV (6.2 eV) increases by about two orders of magnitude (less than two times) from near room temperature to 100 K, and it differs by about two orders of magnitude near room temperature between $h\nu =10.0$ eV and 6.2 eV. These observations are reproduced by our temperature-dependent model $S_E$ calculations (Fig. 2d) with the same set of model parameters as those used in Fig. 2a-2c and two $\alpha_L$ values for the two $h\nu$ values.

## Further experimental tests for SrTiO$_3$

The preceding agreements between theory and experiment suggest that our recycling mechanism can explain the intense coherent photoemission recently observed[6] from SrTiO$_3$. Further in-depth studies guided by our DFT and model calculations would be useful to gain a deeper understanding of this interpretation as follows. The expected value of $\phi_{AATP}$~6.1 eV based on our DFT calculations (Fig. 3b) can be confirmed by measuring the SPS peak intensity with continuously tunable $h\nu$. A systematic $I_0$-dependent study of the SPS peaks will allow a quantitative examination of the superlinear intensity increase (relative to the continuous background) and linewidth narrowing as a function of $I_0$ predicted by our model calculations (Fig. 2a-2b and Extended Data Fig. 1). Beyond these SPS measurements that directly probe the E-band as the photoemission final states, two-photon photoemission spectroscopy[51] can provide access to both the U- and E-bands as the initial states for the probe excitation. The predicted population and effective lifetime inversions between the U- and E-bands in the steady state for high/intermediate $I_0$ (Fig. 2a-2b and Extended Data Fig. 1) as well as their time-dependent evolutions (see Supplementary Information) can, in principle, be verified through the measured $I_0$ and pump-probe time delay dependences.

## Surface effects in SrTiO$_3$

While both of our DFT-based bulk band structure calculations and phenomenological model calculations compare reasonably well with the recent experimental results on SrTiO$_3$(001) (see above)[6], here we turn to consider the role of the SrTiO$_3$ surface. To the extent the 2×1 surface reconstruction is relevant, which appears to be the case in experiments[6], the surface-bulk band alignment is essential to understanding the transport step of the photoemission process, as is the makeup of the surface dipole layer to the escape step[52]. In addition to these well-known surface effects, quantum confinement of the electronic states in the surface layer might also be relevant because it can produce discrete DOS features in the I-band region, see Extended Data Fig. 10. Depending on the nature of the reconstruction, one of these distinct band features could show up at an energy position that exactly satisfies the coincidence condition required for the AATP process, and thus produce a resonant surface effect, which could combine with the bulk AATP process to enhance photoemission from the E-band. Unfortunately, despite intense studies of the 2×1 reconstruction of the SrTiO$_3$ surface, the exact atomic configurations involved remain uncertain[53-55], so that the importance of the issues discussed above for interpreting experiments is currently unclear. However, because of the predominantly bulk nature of the SPS peaks necessitated by the long lifetimes of the associated minority electrons, our picture based on comparisons with the bulk band structure of SrTiO$_3$ provides a viable approach and sets the stage for developing a complete understanding of the related phenomena.

## Materials predictions

To find other material candidates for the recycling mechanism, we used two different materials databases: the Materials Project (https://legacy.materialsproject.org/)[15,16] and the Materials Flatband Database (https://www.topologicalquantumchemistry.fr/flatbands/)[17-20] along with our own database. Because of the limited resolutions used for the calculations of the DOS in these databases, we performed the needed high-resolution calculations for screening potential candidate materials (see Extended Data Table 1 for details related to the materials on display in figures).

In the Materials Project database, we performed a constrained search over all the cubic oxide perovskites with the chemical formula $ABO_3$, where A = Mg, Ca, Sr, and Ba and B = Ti, Zr, Hf, V, Nb, Ta, Cr, Mo, and W. These materials possess band structures similar to $SrTiO_3$. For each material, the E- (or E'-), U-, and L-bands were assigned to the DOS or band-structure features corresponding to those in $SrTiO_3$. The I-band is expected from the coincidence condition to be displaced from the L-band by the amount of the energy separation between the E- (or E'-) and U-bands. Only if the I-band thus determined lies close in energy to a local maximum in the DOS or the saddle point at X in the band structure, was the coincidence condition considered to be satisfied in the material. Selected candidate materials so obtained for supporting the AATP process are displayed in Fig. 4.

In the Materials Flatband Database, we performed an extensive search over all the potential flat-band materials that contain either Kagome, pyrochlore, Lieb, bipartite or split sublattices within the 230 space groups, which involved a total of 28,169 materials[18,20]. An initial screening was first performed to identify materials that show prominent DOS-peak features associated with flat bands in both the low-binding-energy conduction and valence band regions. A quantitative examination of the coincidence condition was then performed on this initial list of materials. For each material, the U- and L-bands were constrained to be one of the most prominent DOS-peak features associated with the flat bands in the low-binding-energy conduction and valence band regions, respectively. The E-band was constrained to be one of the bands that cross Γ. The I-band was located using the coincidence condition from the positions of the E-, U- and L-bands. Only if the energy position of the I-band thus determined was found to be close to that of a local maximum in the DOS or a flat band in the band structure, was the coincidence condition considered to be satisfied in the material. Due to the limited resolution used for the DOS calculations in the database and the excessive amount of data involved over an extended energy window, it is likely that some cases with reasonably met coincidence condition were overlooked in our search process. Selected candidate materials so obtained for supporting the AATP process are displayed in Extended Data Figs. 3 and 4.

**Beyond the traditional Auger processes**

The specific mechanism we propose for the recycling step in this work is distinct from the Auger processes that have been considered in the early and current photoemission theories within either the one- or three-step framework in the following aspects.

*Role of recycling*: While Spicer et al. discussed and modeled the role of Auger processes in their original paper[56], they only considered the process that involves a majority (M) electron recombining with a minority (m) hole and another majority electron being subsequently excited above $E_{vac}$. In this type of traditional Auger process, which we call an m-M-M process, there is no recycling of failed escaping minority electrons. In contrast, our proposed mechanism involves a failed escaping minority (U-band) electron recombining with a minority (L-band) hole to excite a majority (I-band) electron which, we call an m-m-M process. This new process can recycle the energy of the failed escaping minority electron. The difference highlights the unique role of this new AATP process in the photoemission phenomenon.

*Participation of a minority electron in Auger recombination*: This is a defining feature of the AATP process that enables the recycling step, as described above.

*Second-order nature*: Unlike the traditional (m-M-M) Auger process, our novel (m-m-M) Auger process is second order in the minority carrier density that involves minority electrons and holes on the U- and L-bands. Owing to their flat-band nature, both the U- and L-bands have high densities of states, which allows them to be effective in collecting thermalized minority carriers. Therefore, their associated minority carrier densities are not necessarily small (see also below), making possible a meaningful contribution of the AATP process to photoemission with nontrivial consequences revealed by our four-step model calculation.

*Different origin of the minority carriers*: The probability for one photon of energy $h\nu$ to create a minority hole at a specific energy $E_0(\ll h\nu)$ is low as there are in general a great number of possible initial states available for the photoexcitation. However, the probability for the same photon to create a minority hole in the energy range $[E_0, h\nu]$ (for metals) is much higher, and this photoexcited minority hole has a certain probability of lowering its energy to $E_0$ through a later thermalization or scattering process. The traditional Auger recombination only considers minority holes at $E_0$ created in the former case as a result of direct photoexcitation, whereas our novel Auger recombination effectively considers minority holes at $E_0$ created in the latter case as a result of combined photoexcitation and thermalization/scattering. The same situation also applies to minority electrons. The minority electrons and holes thus created are of much higher densities in our case, and are capable of producing much stronger electron emissions than in the traditional Auger case, as needed to explain the recent experimental observations[6] on SrTiO$_3$.

*Discrete nature of the final states of the excited majority electrons*: The relevant final states are the discrete E-band states in the AATP case, as opposed to the continuum in the traditional case.

*Energy specificity and discrete spectrum*: As all the states that are involved (including the flat bands in the conduction and valence band region and the discrete states above $E_{vac}$) have highly specific energy positions, the AATP process can produce a discrete emission spectrum with sharp features. In contrast, the traditional Auger processes considered in current photoemission theories generally involve dispersive, overlapping valence states and the continuum above $E_{vac}$, thereby producing incoherent spectra with broad structures.

In connection with the recent SrTiO$_3$ experiment[6], we note that the relevance of the traditional Auger processes can be entirely ruled out based on a simple argument. In SrTiO$_3$, the conduction band region is separated from the valence band region by an energy gap of ~3.25 eV. The valence band has an overall bandwidth of ~4-5eV, which is separated by another energy gap of ~10 eV from the (Sr) semi-core states that lie below. In a traditional Auger process excited by a low-energy photon, say of energy ~10 eV, the minority hole can only be created in the valence band of SrTiO$_3$, which will recombine with a majority electron also in the valence band. This recombination will release energy $\Delta E$ to another majority electron in the valence band that will be promoted to an unoccupied state of $\Delta E$ higher energy. In this case, the maximum $\Delta E$ is the size of the valence band bandwidth, which is insufficient to promote a valence electron above the vacuum level located at least 2.8 eV above the CBM[6]. Therefore, the traditional Auger process is not active for photoemission from SrTiO$_3$ excited by the low-energy photons used in the recent experiment[6]. In comparison, our AATP process can occur above a much lower photon energy threshold $\phi_{AATP}$~6.1 eV, compatible with the experiment. This adds another distinction to our AATP process.

## Beyond the one-step model

As we have pointed out, the AATP process can be conceived as a novel type of Auger or electron-electron scattering process. Such a process is not captured within the current one-step photoemission framework[8-11,57,58], which is built on a one-particle formalism[8] and mainly accounts for the primary (photoexcited) electrons. Effects of scattering with other degrees of freedom, including the secondary electrons and minority holes, are introduced by adding an ad hoc self-energy term in the initial- and final-state Green's functions[11,57]. These effects, which are essential to the Auger emissions, in principle require descriptions on the many-body level[59-64]. Such theories, however, either are numerically intractable[59,60] or rely on rudimentary treatments of correlation effects (not at the *ab initio* level)[61-64].

Including the traditional Auger processes in the one-step model based on a single-particle approximation would be difficult but it will not be impossible. However, including our novel AATP process will require a major overhaul of the existing single-particle approximations within the one-step framework.

We note that the current approach to resonant photoemission[65] is to model a specific form of the traditional Auger process that only involves the participation of a majority electron in the Auger recombination with the minority hole generated by a preceding photoexcitation process. As mentioned previously, the AATP process differs from such a traditional Auger process in many aspects. In particular, the AATP process involves the participation of a minority electron in the Auger recombination with a minority hole, both of which are generated by the combined photoexcitation and thermalization processes. These distinctive features of the AATP process are absent in the existing formalism for resonant photoemission.

It should also be noted that the existing treatment of time-dependent photoemission[66,67] involves modeling the photoexcitation process by a probe pulse of a system with a population of minority carriers excited by a preceding pump pulse. Relaxation dynamics of the minority carriers is assumed to lie within a restricted energy range (much smaller than the work function) above $E_F$ and a short time duration (sub-picosecond) after the pump excitation, and it only involves electronic degrees of freedom. Such a relaxation, however, is far from being a full-fledged version of thermalization that is required for the AATP process. Recent progress[68] has been made to include relaxation via phonons, interactions with impurities, and Coulomb scattering, but only for temporally separated pump excitations, whereas our experimentally relevant AATP process is driven by a continuous-wave excitation. Crucially, current models for pump-probe photoemission lack a basic Auger component required for modelling the AATP process.

Notably, a model[69] recently introduced for pump-probe Auger electron spectroscopy that involves a non-traditional (m-m-M) Auger process may be more promising. In that case, nevertheless, minority valence electrons and core holes are created at separate times within the pump and probe durations (femtoseconds), respectively. Electronic relaxation was considered only for the minority electrons (not for the core holes) within a restricted energy range and small time duration as in the case of pump-probe photoemission[66,67], although a certain form of the core-hole relaxation[70] might in principle be added to the model. In light of the large separation of relevant time scales (from femtoseconds to nanoseconds), treating the AATP process driven by adequately thermalized minority electrons and holes in the presence of a continuous excitation remains a formidable task, even at the many-body level, and it will be very challenging to cast it in a computationally tractable single-particle-approximation scheme.

**Data availability**

The data that support the findings of this study are available from the corresponding authors upon reasonable request.

**Code availability**

The computer codes that support the findings of this study are available from the corresponding authors upon reasonable request.

42. Kresse, G. & Joubert, D. From ultrasoft pseudopotentials to the projector augmented-wave method. *Phys. Rev. B* **59**, 1758 (1999).
43. Kresse, G. & Hafner, J. Ab initio molecular dynamics for liquid metals. *Phys. Rev. B* **47**, 558 (1993).
44. Perdew, J. P., Burke, K. & Ernzerhof, M. Generalized gradient approximation made simple. *Phys. Rev. Lett.* **77**, 3865-3868 (1996).
45. Wahl, R., Vogtenhuber, D. & Kresse, G. $SrTiO_3$ and $BaTiO_3$ revisited using the projector augmented wave method: Performance of hybrid and semilocal functionals. *Phys. Rev. B* **78**, 104116 (2008).
46. Sun, J., Ruzsinszky, A. & Perdew, J. P. Strongly constrained and appropriately normed semilocal density functional. *Phys. Rev. Lett.* **115**, 036402 (2015).
47. Marzari, N. & Vanderbilt, D. Maximally localized generalized Wannier functions for composite energy bands. *Phys. Rev. B* **56**, 12847–12865 (1997).
48. Mostofi, A. A. et al. wannier90: A tool for obtaining maximally-localised Wannier functions. *Comput. Phys. Commun.* **178**, 685 (2008).
49. Sancho, M. P. L., Sancho, J. M. L. & Rubio, J. Highly convergent schemes for the calculation of bulk and surface Green functions. *J. Phys. F: Met. Phys.* **15**, 851 (1985).
50. van Benthem, K., Elsässer, C. & French, R. H. Bulk electronic structure of $SrTiO_3$: Experiment and theory. *J. Appl. Phys.* **90**, 6156-6164 (2001).
51. Sobota, J. A. et al. Direct optical coupling to an unoccupied Dirac surface state in the topological insulator $Bi_2Se_3$. *Phys. Rev. Lett.* **111**, 136802 (2013).
52. Martinelli, R. U. & Fisher, D. G. The application of semiconductors with negative electron affinity surfaces to electron emission devices. *Proc. IEEE* **62**, 1339-1360 (1974).
53. Jiang Q. D. & Zegenhagen, J. (6×2) and c(4×2) reconstruction of $SrTiO_3(001)$. *Surf. Sci.* **425**, 343–354 (1999).

54. Erdman, N. et al. The structure and chemistry of the TiO$_2$-rich surface of SrTiO$_3$ (001). *Nature* **419**, 55 (2002).
55. Becerra-Toledo, A. E., Castell, M. R. & Marks, L.D. Water adsorption on SrTiO$_3$(001): I. Experimental and simulated STM. *Surf. Sci.* **606**, 762–765 (2012).
56. Berglund, C. N. & Spicer, W. E. Photoemission Studies of Copper and Silver: Theory. *Phys. Rev.* **136**, A1030–A1044 (1964).
57. Minár, J., Braun, J., Mankovsky, S. & Ebert, H. Calculation of angle-resolved photo emission spectra within the one-step model of photo emission—Recent developments. *J. Electron Spectrosc. Relat. Phenom.* **184**, 91-99 (2011).
58. Getzlaff, M., Bansmann, J., Braun, J. & Schönhense, G. Surface magnetism of iron on W(110). *Z. Phys. B* **104**, 11-20 (1997).
59. Schaich, W. L., Ashcroft, N. W. Model calculations in the theory of photoemission. *Phys. Rev. B* **3**, 2452 (1971).
60. Caroli, C., Lederer-Rozenblatt, D., Roulet, B. & Saint-James, D. Inelastic effects in photoemission: Microscopic formulation and qualitative discussion. *Phys. Rev. B* **8**, 4552 (1973).
61. Cini, M. Two hole resonances in the XVV Auger spectra of solids. *Solid State Commun.* **24**, 681-684 (1977).
62. Sawatzky, G. A. Quasiatomic Auger Spectra in Narrow-Band Metals. *Phys. Rev. Lett.* **39**, 504-507 (1977).
63. Gunnarsson, O. & Schönhammer, K. Dynamical theory of Auger processes. *Phys. Rev. B* **22**, 3710-3733 (1980).
64. Verdozzi, C., Cini, M. & Marini, A. Auger spectroscopy of strongly correlated systems: present status and future trends. *J. Electron Spectrosc. Relat. Phenom.* **117-118**, 41-55 (2001).
65. Da Pieve, F. & Krüger, P. Real-space Green's function approach to angle-resolved resonant photoemission: Spin polarization and circular dichroism in itinerant magnets. *Phys. Rev. B* **88**, 115121 (2013).
66. Braun, J., Rausch, R., Potthoff, M., Minár, J. & Ebert, H. One-step theory of pump-probe photoemission. *Phys. Rev. B* **91**, 035119 (2015).
67. Freericks, J. K., Krishnamurthy, H. R. & Pruschke, Th. Erratum: Theoretical Description of time-Resolved Photoemission Spectroscopy: Application to Pump-Probe Experiments [Phys. Rev. Lett. 102, 136401 (2009)]. *Phys. Rev. Lett.* **119**, 189903 (2017).
68. Kemper, A. F., Abdurazakov, O. & Freericks, J. K. General Principles for the Nonequilibrium Relaxation of Populations in Quantum Materials. *Phys. Rev. X* **8**, 041009 (2018).
69. Rausch, R. & Potthoff, M. Pump-probe Auger-electron spectroscopy of Mott insulators. *Phys. Rev. B* **99**, 205108 (2019).
70. Nocera, A. & Feiguin, A. Auger spectroscopy beyond the ultra-short core-hole relaxation time approximation. Preprint at https://doi.org/10.48550/arXiv.2304.15001 (2023).

**Acknowledgements**
We thank B. Singh and Taofei Zhou for discussion and K. Tanaka for experimental support. The work at Northeastern University was supported by the Air Force Office of Scientific Research under Award No. FA9550-20-1-0322 and benefited from the computational resources of Northeastern University's Advanced Scientific Computation Center and the Discovery Cluster. The work at Westlake University was supported by the National Natural Science Foundation of China (Grants No. 12274353, 11874053), Zhejiang Provincial Natural Science Foundation of China (LZ19A040001), Yecao Tech. Co. (2021ORECH020015), and the Westlake Instrumentation and Service Center for Physical Sciences.

**Author contributions**
A.B. and R.-H.H. conceived the project, secured funding, and guided the investigations. R.-H.H., R.S.M., M.M., W.-C.C., and B.B. contributed to the revelation of the AATP process. R.-H.H. proposed the conceptual idea of the recycling step and a simplified version of the phenomenological model. R.S.M. developed and, together with M.M., solved the phenomenological model in the full version. M.M., W.-C.C. and B.G. performed DFT and Wannier orbital calculations, respectively, under the guidance of R.-H.H., R.S.M., B.B. and A.B.. C.H. and P.R. performed related experiments, data analysis and comparison with the calculations, under the guidance of C.Z. and S.L.. R.-H.H., R.S.M., M.M., C.H. and W.-C.C. contributed to the manuscript writing with input from A.B. and B.B.. All authors contributed to discussions.

**Competing interests**
The authors declare no competing interests.

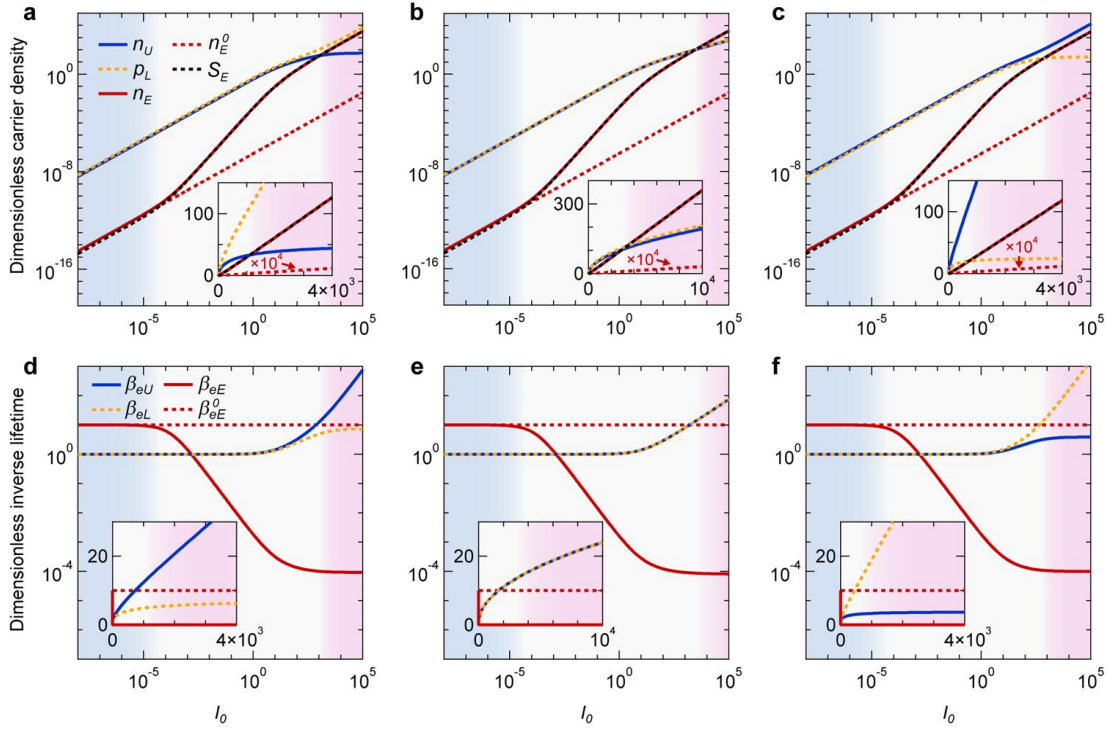

**Extended Data Fig. 1 | Representative steady-state solutions of the phenomenological three-band model in the three parameter-space regimes. a, b, c,** Carrier density, $n_i$, and **d, e, f,** effective inverse lifetimes, $\beta_{ei}$, for the E-, U- and L-bands as a function of $I_0$ for the regimes of $\gamma'_L > 0$ (**a, d**), $\gamma'_L = 0$ (**b, e**), and $\gamma'_L < 0$ (**c, f**), where $\gamma'_L$ is a function of several model parameters, see Methods for definitions of parameters. Main panels (insets) are on logarithmic (linear) scales. $n_E^0$ in the insets of **a, b, c** is magnified by 10,000 times. In each regime, different $I_0$-regions in the main panels and the insets are shaded in different colors. For high $I_0$ values, $n_E$ is always linear in $I_0$ with a parameter-dependent slope; $n_U$ and $p_L$ have distinct behaviors in the three regimes: $n_U$ ($p_L$) saturates to a finite level, while $p_L$ ($n_U$) is linear in $I_0$ for $\gamma'_L > 0$ ($\gamma'_L < 0$); both $n_U$ and $p_L$ are proportional to $\sqrt{I_0}$ with a parameter-dependent prefactor for $\gamma'_L = 0$, see Methods for details. **b, e** are reproduced from Fig. 2a-2b. Other panels share the same values of model parameters except for $\alpha_U$ and $\alpha_L$: $\alpha_U = 0.4, \alpha_L = 0.5$ for **a, d** ($\gamma'_L = 0.067$); $\alpha_U = 0.4625, \alpha_L = 0.5$ for **a, d** ($\gamma'_L = 0$); $\alpha_U = 0.5, \alpha_L = 0.4$ for **a, d** ($\gamma'_L = -0.141$), which are marked in Fig. 2c in black diamond, pentagon and square symbols, respectively.

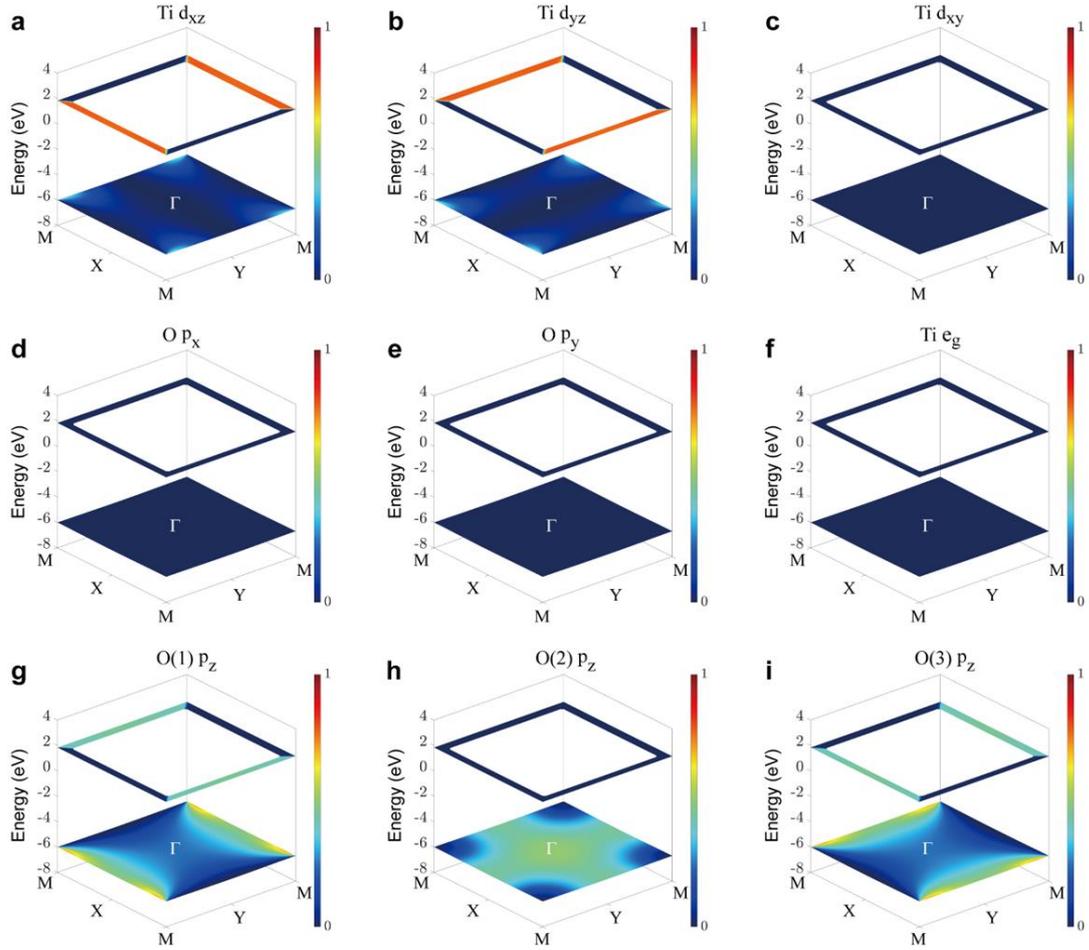

**Extended Data Fig. 2 | Orbital character analysis for the U- and L-bands in SrTiO₃.**
**a-i,** Weight distributions for indicated atomic orbital components in 2D momentum space ($k_x$-$k_y$ plane) at the energy positions of the U- and L-bands. Atomic orbitals are used in the calculated Wannier wave functions (see Methods) as its basis, and the individual orbital weight is the decomposition coefficient squared of the wave function for each orbital basis. For visualization purposes, the orbital weights have been magnified by 2 times for the Ti 3d orbitals of L-band and for the O 2p orbitals of U-band.

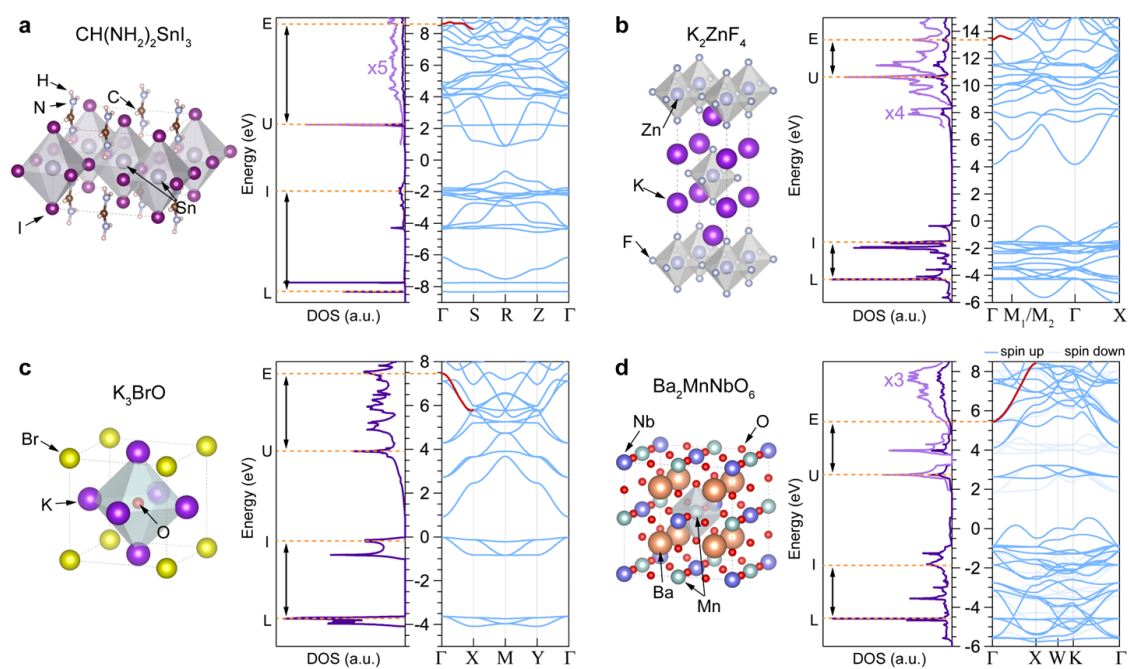

**Extended Data Fig. 3 | Materials predictions of the perovskite-derivative candidates for the recycling mechanism. a-c,** Crystal structures, calculated DOSs and band structures of $CH(NH_2)_2SnI_3$ (**a**), which is a hybrid halide perovskite known for making a perovskite solar cell with 9 % efficiency[71], $K_2ZnF_4$ (**b**), which is a Ruddlesden-Popper layered fluoride perovskite of $La_2CuO_4$-type structure[72], $K_3BrO$ (**c**), which is an anti-perovskite halide oxide[73], and $Ba_2MnNbO_6$ (**d**), which is magnetic double perovskite[74]. For each material, the U- and L-bands are constrained to be one of the most prominent DOS peak features associated with flat bands in the low-binding-energy conduction and valence band regions, respectively. The E-band is constrained to be one of the bands that cross Γ. The I-band is obtained using the coincidence condition, and it is displaced from the L-band by the amount of the energy separation between the E- and U-bands, as marked by the arrows. Energy positions of the E-, U-, I- and L-bands thus determined are marked by the orange dashed lines. A portion of the E-band is highlighted in red. Note that the I-band in each case lies in energy close to a local maximum in the DOS or a flat band in the band structure, which indicates that the coincidence condition is met. Lattice parameters used in the bulk calculations here are taken from the databases[15-20] and summarized in Methods. For **d**, calculation yields a ferromagnetic ground state with spin-polarized band structure around $E_F$ (bands in the two spin channels shown in blue and light blue, respectively). The purple curve in the DOS shown in **a** (**b**) is a blow-up of the conduction-band portion of the blue curve by 5 (4) times.

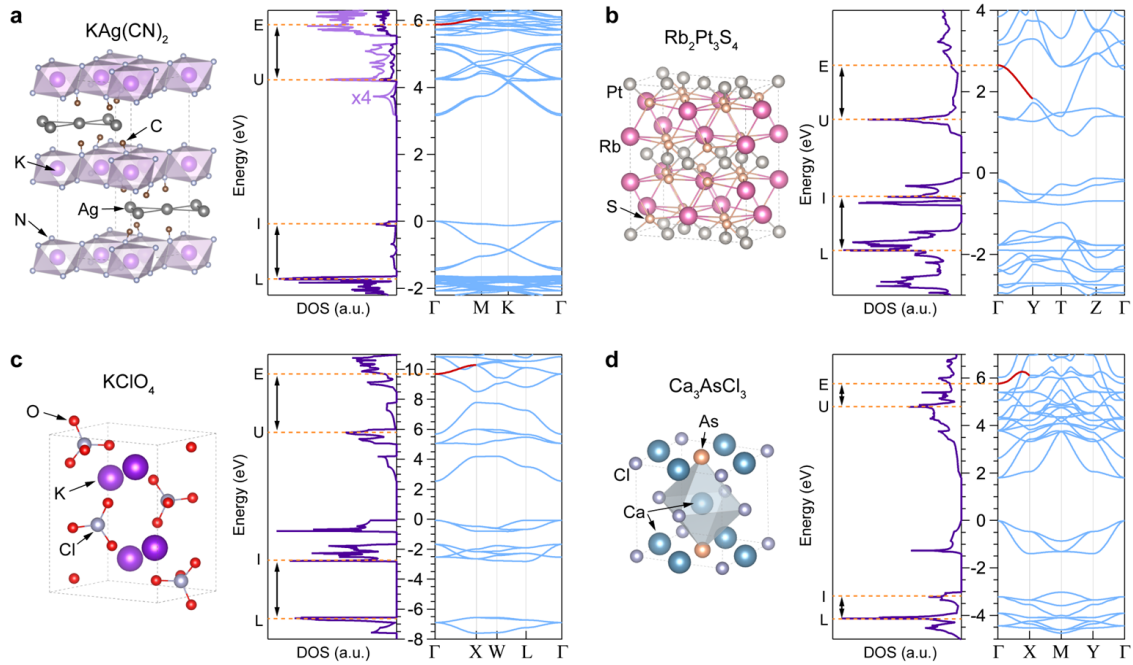

**Extended Data Fig. 4 | Materials predictions of the non-perovskite candidates for the recycling mechanism. a-d,** Crystal structures, calculated DOSs and band structures of KAg(CN)$_2$ (**a**), which hosts a 2D Kagome sublattice formed by the Ag atoms (in silver) and features a topological flat band located near $E_F$ in the band structure[18], Rb$_2$Pt$_3$S$_4$ (**b**), which has a layered structure with a Kagome sublattice formed by the Pt atoms (in silver)[75], KClO$_4$ (**c**), which has a layered cubic structure[76], and Ca$_3$AsCl$_3$ (**d**), which is a mixed-anion compound of a rock-salt structure with ordered cation vacancies[77]. For each material, the U- and L-bands are constrained to be one of the most prominent DOS peak features associated with flat bands present in the low-binding-energy conduction and valence band regions, respectively. The E-band is constrained to be one of the bands that cross Γ. The I-band is obtained using the coincidence condition, and it is displaced from the L-band by the amount of the energy separation between the E- and U-bands, as marked by the arrows. Energy positions of the E-, U-, I- and L-bands thus determined are marked by the orange dashed lines. A portion of the E-band is highlighted in red. Note that the I-band in each case lies in energy close to a local maximum in the DOS or a flat band in the band structure, which indicates that the coincidence condition is met. Lattice parameters used in the bulk calculations here are taken from the database[17-20] and summarized in Methods. The purple curve in the DOS shown in **a** is a blow-up of the conduction-band portion of the blue curve by 4 times.

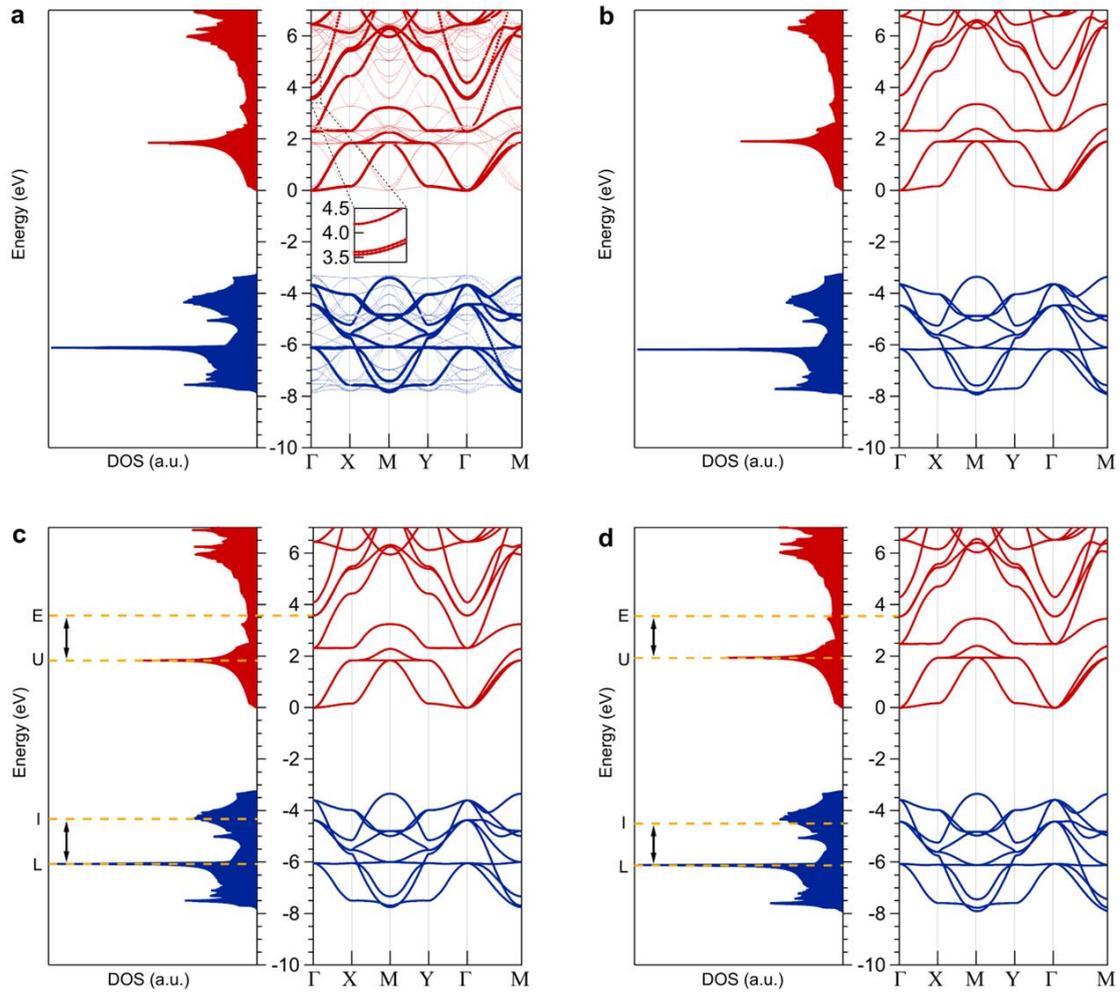

**Extended Data Fig. 5 | Bulk band structures of SrTiO$_3$ calculated in different cases.** DOS (left) and band structure (right) in the tetragonal (**a**) and cubic (**b-d**) phases. Results were obtained using the PBE (**a, c, d**) and SCAN (**b**) functionals. Lattice parameters were relaxed for **b** (3.908 Å) and **c** (3.941 Å), but not for **a** (5.563 Å and 7.900 Å) and **d** (3.900 Å). See Methods for related descriptions. In **c** and **d**, the E-, U-, and L-bands are assigned to similar band-structure features as in Fig. 3a. The I-band is obtained using the coincidence condition, and it is displaced from the L-band by the amount of the energy separation between the E- and U-bands, as marked by the arrows. Energy positions of the E-, U-, I- and L-bands thus determined are marked by the orange dashed lines. Going from **c** to **d**, an increased misalignment in energy between the I-band and the local maximum in the DOS near -4.5 eV suggests that the lattice shrinkage causes the coincidence condition to deteriorate. **c** is reproduced from Fig. 3a-3b.

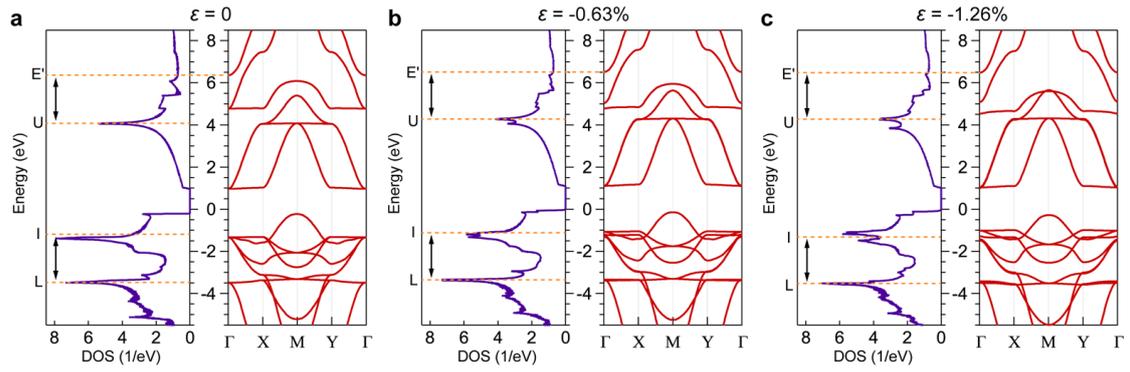

**Extended Data Fig. 6 | Effect of strain on the coincidence condition in WO$_3$. a-c,** Calculated DOSs and band structures of WO$_3$ at different in-plane compressive strain levels ($\varepsilon < 0$) as labelled. Strain is applied by maintaining the unit-cell volume to the original lattice parameters (3.863 Å) taken from the database[15,16]. In each case, the E'-, U-, and L-bands are assigned to the DOS or band-structure features corresponding to those in SrTiO$_3$ (Fig. 3a). The I-band is obtained using the coincidence condition, and it is displaced from the L-band by the amount of the energy separation between the E'- and U-bands, as marked by the arrows. Energy positions of the E'-, U-, I- and L-bands thus determined are marked by the orange dashed lines. Note that the I-band lies in energy close to a local maximum in the DOS in **b**, which indicates that the coincidence condition is best met for $\varepsilon = -0.63$ %. **b** is reproduced from Fig. 4e.

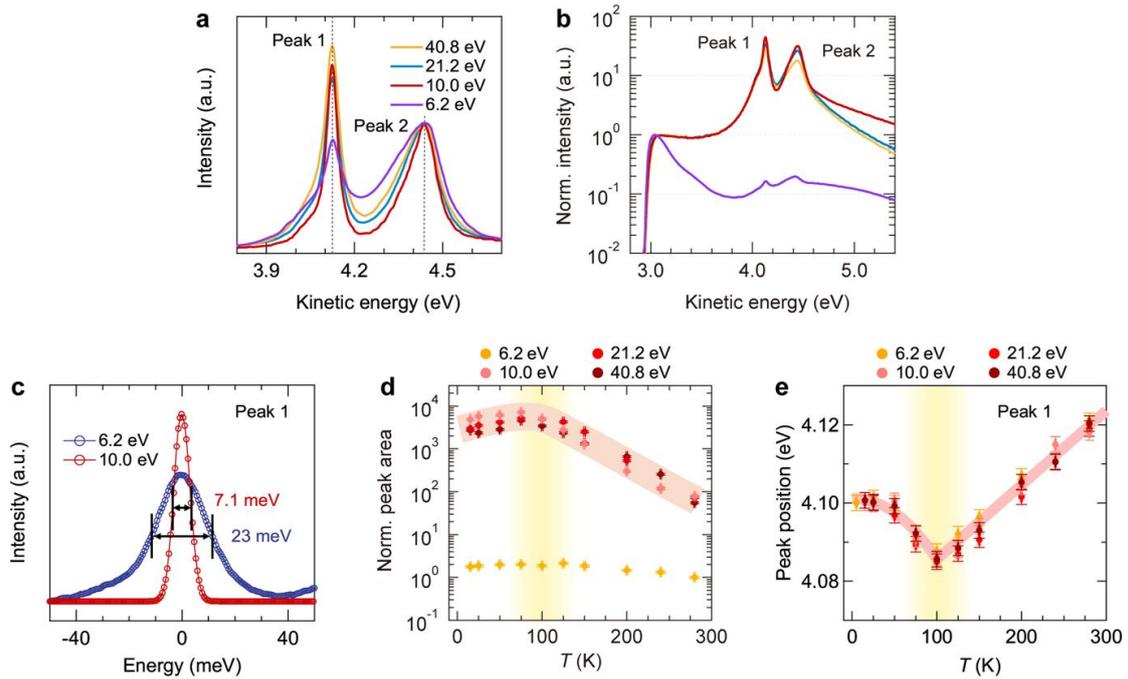

**Extended Data Fig. 7| Experimental results of the SPS measurements on SrTiO$_3$(001). a,** SPS measured at normal emission near room temperature with different $h\nu$ values as indicated and plotted on a linear scale with intensity normalized to Peak 2. **b,** SPS plotted on a logarithmic scale with intensity normalized to the work-function cutoff. Angle integration of the measured SPS over ±7.5° about the surface normal direction was performed on the same data set shown in **a** for the comparison of the momentum-integrated intensity around $k_\parallel$ =0. **c,** SPS measured at normal emission and 5 K centered at Peak 1 measured with $h\nu = 6.2$ eV and 10.0 eV that shows a difference in the energy linewidth as marked. The SPS intensity is rescaled for ease of comparison. Note that the effective experimental resolution is estimated to be somewhere between 4 meV and 7.1 meV. This can possibly make the apparent linewidth much larger than the intrinsic one for $h\nu = 10.0$ eV. **d, e**, Temperature dependence of the integrated intensity and energy position of Peak 1, respectively, measured with different $h\nu$ values as indicated. Intensity integration was performed in angle over ±7.5° about the surface normal direction and in energy over Peak 1 to extract the peak area. Integrated intensity (peak area) was obtained for each $h\nu$ is normalized to the room-temperature value for $h\nu = 6.2$ eV according to the respective ratio determined from **b**. Vertical shaded areas mark the same temperature window across which trend changes are observed. Pink ribbons are guides to the eye; the one in **e** is obtained from a simple curve fitting (see Methods). All results (and error bars) are reproduced or adapted from Ref. 6.

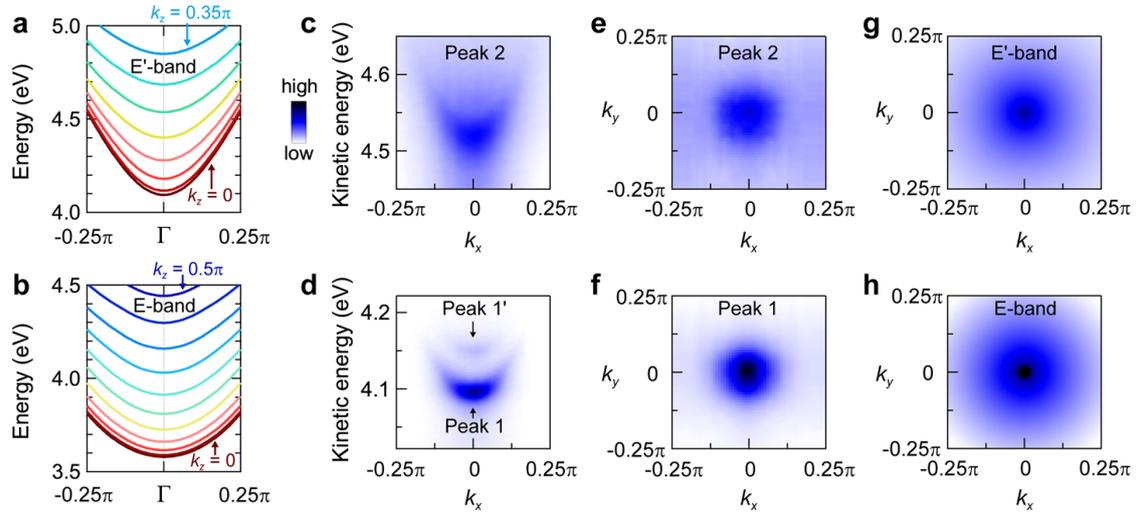

**Extended Data Fig. 8 | Comparison between the E-band (E′-band) in DFT calculations and Peak 1 (Peak 2) in experiments. a, b,** $k_z$-dependent dispersions of the E′- and E-bands, respectively. The range of $k_z$ variation (in $0.05\pi$ steps) in the relevant energy window is marked in each case. **c, d,** Band dispersions associated with Peaks 2 and 1 (and 1′), respectively. **e, f, (g, h,)** Dispersion contours associated with Peaks 2 and 1 (the E′- and E-bands) near their respective band bottoms at Γ. The experimental results in **c, d, e, f** are reproduced from Ref. 6, which were obtained at 5 K with $h\nu$ =6.2 eV.

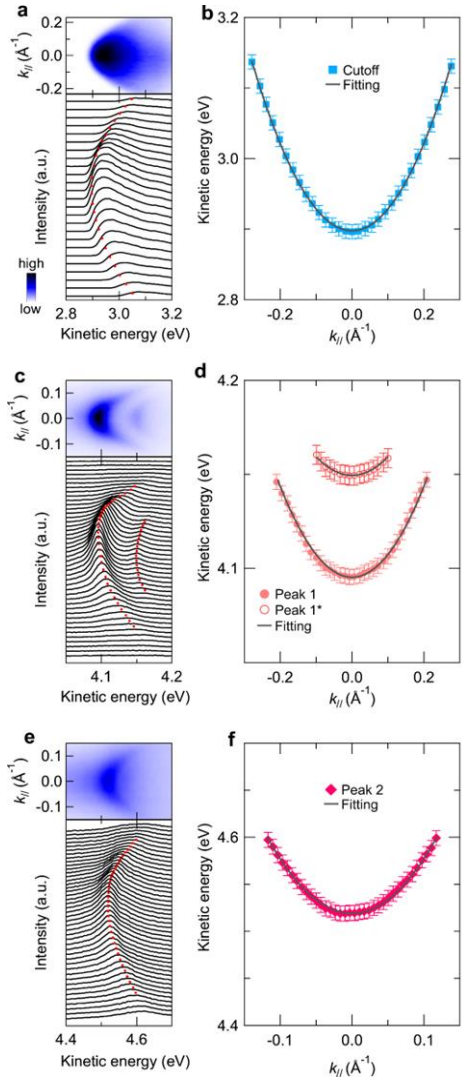

**Extended Data Fig. 9 | Effective mass analysis for the experiment on SrTiO$_3$(001). a,** SPS image plot and SPS measured at 5 K with $h\nu$ =6.2 eV in the energy region of the work-function cutoff. **b,** Extracted band dispersion for the cutoff feature as traced in **a**. **c, d, (e, f,)** Related results obtained in the Peak 1 (Peak 2) region. Curve fitting with a simple parabola to all the extracted dispersions yields the following effective masses for bands associated with Peaks 1, 1' and 2, respectively: $m_{Peak1} \sim 0.64 \pm 0.06 m_e$, $m_{Peak1'} \sim 0.78 \pm 0.17 m_e$, $m_{Peak2} \sim 0.53 \pm 0.07 m_e$, where $m_e$ is the free electron mass determined by a similar fit to the cutoff dispersion in **b**. Error bars reflect the uncertainty in determining the energy position of the related feature subject to fitting. Raw data were obtained from Ref. 6.

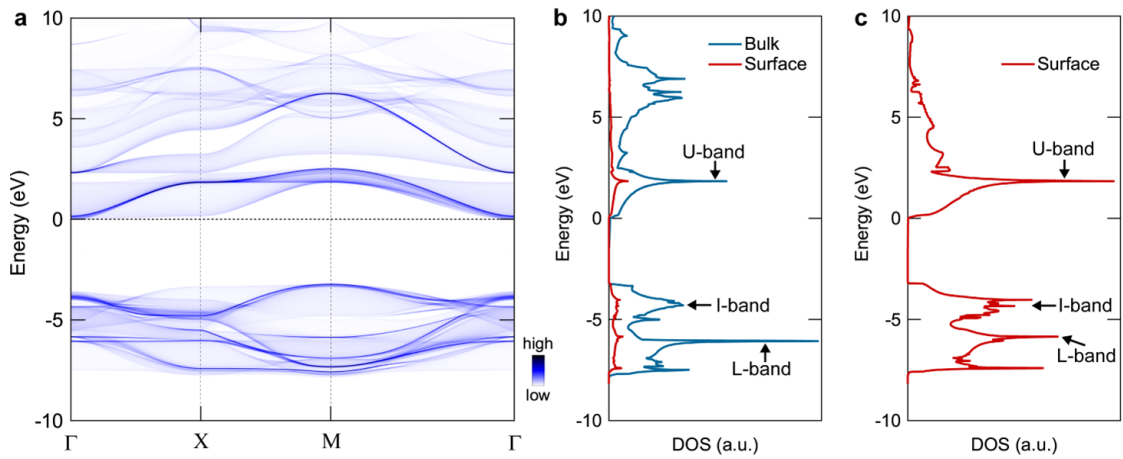

**Extended Data Fig. 10 | Surface and bulk band structures of a semi-infinite SrTiO₃(001) slab. a,** Band structure of the SrTiO₃(001) surface with TiO₂ termination (see Methods). Spectral weight of the surface TiO₂ layer is shown in color scale. **b, c,** DOS of the same surface in comparison with the bulk DOS in **b**. The surface/bulk DOS is calculated by integrating the surface/bulk spectral weight over the surface/bulk Brillouin zone. Distinct surface-state features appear in the I-band region that are absent in the bulk states and have an enhanced DOS relative to those of the U- and L-bands.

| Cubic oxide perovskites | Space group | Wannier orbitals | DOS k-mesh | DOS method | Database source |
|---|---|---|---|---|---|
| BaTiO$_3$ | 221 | Ba $s, p, d$, Ti $s, p, d$, O $s, p$ | 501×501×501 | No smearing | MP |
| KTaO$_3$ | 221 | K $s, p, d$, Ta $s, p, d$, O $s, p$ | 501×501×501 | No smearing | MP |
| BaNbO$_3$ | 221 | Ba $s, p, d$, Nb $s, p, d$, O $s, p$ | 501×501×501 | No smearing | MP |
| WO$_3$ | 221 | W $s, p, d$, O $s, p, d$ | 501×501×501 | No smearing | MP |
| SrVO$_3$ | 221 | Sr $s, p, d$, V $s, p, d$, O $s, p$ | 501×501×501 | No smearing | MP |
| **Perovskite derivatives** | | | | | |
| CH(NH$_2$)$_2$SnI$_3$ | 38 | C $p$, H $s$, N $p$, Sn $s, p, d$, I $s, p, d$ | 51×51×51 | Adaptive smearing | FD |
| K$_2$ZnF$_4$ | 139 | K $p, d$, Zn $s, p, d$, F $p$ | 51×51×51 | Adaptive smearing | FD |
| K$_3$BrO | 221 | K $s, p, d$, Br $s, p, d$, O $s, p$ | 501×501×501 | No smearing | FD |
| Ba$_2$MnNbO$_6$ | 225 | Ba $s, p, d$, Mn $s, p, d$, Nb $s, p, d$, O $s, p$ | 51×51×51 | Adaptive smearing | MP |
| **Non-Perovskites** | | | | | |
| KAg(CN)$_2$ | 163 | K $s, p, d$, Ag $s, p, d$, C $s, p$, N $s, p$ | 101×101×41 | Adaptive smearing | FD |
| Rb$_2$Pt$_3$S$_4$ | 69 | Rb $d$, Pt $s, p, d$, S $p$ | 51×51×51 | Adaptive smearing | FD |
| KClO$_4$ | 216 | K $d$, Cl $s, p$, O $s, p$ | 51×51×51 | Adaptive smearing | FD |
| Ca$_3$AsCl$_3$ | 221 | Ca $s, p, d$, As $p$, Cl $p$ | 501×501×501 | No smearing | FD |

**Extended Data Table 1 | DOS calculations for the material candidates for the recycling mechanism.** The orbitals used to generate the Wannier function are shown for each candidate material that belongs to the cubic oxide perovskites (first section), the perovskite derivatives (second section), or the non-perovskites (third section) in the space group as specified. For each material, band structure is calculated within the primitive unit cell; the size of the $k$-mesh used for the first Brillouin zone and the method used to calculate the DOS are specified. The adaptive smearing method was implemented in the Wannier90 suite of codes[78]. The lattice parameters used in the calculation were taken either from the (legacy) Materials Project (MP)[15,16] or the Materials Flatband Database (FD)[17-20].

71. Shao, S. et al. Highly reproducible Sn-based hybrid perovskite solar cells with 9% efficiency. *Adv. Energy Mater.* **8**, 1702019 (2018).
72. Rauh, H., Geick, R., Lehner, N., Bouillot, J. & Eckold, G. Phonon dispersion in the perovskite-type layer material $K_2ZnF_4$. *Phys. Status Solidi B* **115**, 463-470 (1983).
73. Jansen, M. The chemistry of gold as an anion. *Chem. Soc. Rev.* **37**, 1826-1835 (2008).
74. Rama, N. et al. Study of magnetic properties of $A_2B'NbO_6$ (A=Ba, Sr, BaSr; and B'=Fe and Mn) double perovskites. *J. Appl. Phys.* **95**, 7528-7530 (2004).
75. Oh, G. N., Choi, E. S. & Ibers, J. A. Syntheses and characterization of nine quaternary uranium chalcogenides among the compounds $A_2M_3UQ_6$ (A = K, Rb, Cs; M = Pd, Pt; Q = S, Se). *Inorg.* **51**, 4224-4230 (2012).
76. Raghurama, G., Al-Dhahir, T. A. & Bhat, H. L. Optical studies on the orthorhombic-cubic transitions for $KClO_4$, $RbClO_4$ and $CsClO_4$. *J. Phys. C: Solid State Phys.* **20**, 4505-4511 (1987).
77. Schäfer, H. On the problem of polar intermetallic compounds: the stimulation of E. Zintl's work for the modern chemistry of intermetallics. *Ann. Rev. Mater. Sci.* **15**, 1-42 (1985).
78. Yates, J. R., Wang, X., Vanderbilt, D. & Souza, I. Spectral and Fermi surface properties from Wannier interpolation. *Phys. Rev. B* **75**, 195121 (2007).